\documentclass[showpacs,prb,superscriptaddress,twocolumn]{revtex4-1}
\usepackage[table]{xcolor}
\usepackage{textcomp}
\usepackage{amsmath}
\usepackage{amssymb}
\usepackage{}
\usepackage{graphicx} 
\usepackage{}
\usepackage{pgf}
\usepackage{hyperref}
\usepackage{epstopdf}

\newcommand{\ket}[1]{\left|#1\right\rangle}
\newcommand{\bra}[1]{\left\langle #1\right|}
\newcommand{\ud}{\mathrm{d}}

\newcommand{\mean}[1]{\left\langle #1\right\rangle}

\newcommand{\nep}{\textrm{e}}
\newcommand{\atan}{\operatorname{atan}}
\newcommand{\sign}{\operatorname{sign}}

\newcommand{\opbfgamma}[1]{{\widehat{\boldsymbol{\gamma}}^{\phantom \dagger}}_{#1}}
\newcommand{\opbfgammadag}[1]{{\widehat{\boldsymbol{\gamma}}^{\dagger}}_{#1}}

\newcommand{\opbfc}[1]{{\widehat{\boldsymbol{c}}^{\phantom \dagger}}_{#1}}
\newcommand{\opbfcdag}[1]{{\widehat{\boldsymbol{c}}^{\dagger}}_{#1}}

\newcommand{\Real}{\operatorname{\Re{\rm e}}}
\newcommand{\Aimag}{\operatorname{\Im{\rm m}}}
\newcommand{\opgamma}[1]{{\hat{\gamma}^{\phantom \dagger}}_{#1}}
\newcommand{\opgammadag}[1]{{\hat{\gamma}^{\dagger}}_{#1}}
\newcommand{\opc}[1]{{\hat{c}^{\phantom \dagger}}_{#1}}
\newcommand{\opcdag}[1]{{\hat{c}^{\dagger}}_{#1}}
\newcommand{\opb}[1]{{\hat{b}^{\phantom \dagger}}_{#1}}
\newcommand{\opbdag}[1]{{\hat{b}^{\dagger}}_{#1}}

\def\ocite#1{{[\onlinecite{#1}]}}

\begin{document}

\title{Spin and topological order in a periodically driven spin chain.}

\author{Angelo Russomanno}
\affiliation{NEST, Scuola Normale Superiore \& Istituto Nanoscienze-CNR, I-56126 Pisa, Italy}
\affiliation{Abdus Salam ICTP, Strada Costiera 11, I-34151 Trieste, Italy}

\author{Bat-el Friedman}
\affiliation{Department of Physics, Bar-Ilan University, 52900, Ramat Gan Israel}

\author{Emanuele G.~Dalla Torre}
\affiliation{Department of Physics, Bar-Ilan University, 52900, Ramat Gan Israel}

\begin{abstract}
The periodically driven quantum Ising chain has recently attracted a large attention in the context of Floquet engineering. In addition to the common paramagnet and ferromagnet, this driven model can give rise to new topological phases. In this work we systematically explore its quantum phase diagram, by examining the properties of its {\em Floquet ground state}. We specifically focus on driving protocols with time-reversal invariant points, and demonstrate the existence of an infinite number of distinct phases. These phases are separated by second-order quantum phase transitions, accompanied by continuous changes of local and string order parameters, as well as sudden changes of a topological winding number and of the number of protected edge states.  When one of these phase transitions is adiabatically crossed, the correlator associated to the order parameter is nonvanishing over a length scale which shows a Kibble-Zurek scaling. %\textcolor{black}{In some of the phases we find, the Floquet ground state shows time-translation symmetry breaking.} 
\textcolor{black}{In some phases, the Floquet ground state spontaneously breaks the discrete time-translation symmetry of the Hamiltonian.} Our findings provide a better understanding of topological phases in periodically driven clean integrable models.

\end{abstract}

\pacs{}

\maketitle

\section{Introduction}
Atomic-physics experiments were recently able to realize exotic many-body phases that simulate interesting states of condensed-matter systems~\cite{Schneider_RPP12,Houck_Nat,Labrutt:preprint2015,Bloch_Nat}. In contrast to solid-state devices, whose parameters are determined by the properties of the material used for the fabrication, atomic quantum simulators are highly tunable. For example, by tuning the intensity of a laser it is possible to change the depth of the periodic lattice in which ultracold atoms move, and by applying a magnetic field it is possible to control the interactions among them~\cite{Bloch_RMP08}. Some remarkable examples are the realization of the Bose-Hubbard model and its superfluid--Mott-insulator transition~\cite{Greiner_Nat02}, the study of the BCS-BEC crossover~\cite{altm_prl}, the realization of artificial gauge fields~\cite{Dalib_RMP11}, the realization of a fermionic Mott insulator~\cite{jordo_nat08,schne_sci08}, and the experimental detection of Anderson localization~\cite{Roati_Nat2008} and many-body localization~\cite{schreiber2015observation} with ultracold atoms.

Ultracold atoms additionally possess an exquisite isolation from the environment, which enables them to preserve quantum coherence for long times. Thus, these systems are a natural playground to observe nonequilibrium effects associated with time-dependent many-body Hamiltonians. Examples of recent studies that explored the response to  external time-dependent perturbations are given in Refs.~\ocite{Greiner_Nat02bis,Kinoshita_Nat06,Bloch_Nat,Bloch_RMP08}. Two common protocols are the sudden quench (the evolution of the system following an abrupt change of the Hamiltonian) and the linear ramping (the evolution of the system with a parameter varying linearly in time). In particular, if such a parameter is slowly varied across a second order quantum phase transition, many observables show a universal Kibble-Zurek scaling (See Refs.~\ocite{Polkovnikov_RMP11,giamarchi2016strongly,Dziarmaga_AP10} for an introduction).

Here we focus on a third type of time-dependent Hamiltonians, namely periodic driving. Previous studies focused in particular on the relaxation to an asymptotic state and its possible thermal properties~\cite{Kollath_PRL06,Ponte_PRL15,Ponte_AP15,
Kim_PRE14,Polkovnikov_NatPhys11,Emanuele_2014:preprint,Dalessio_AP13,Rigol_PRX14,Russomanno_PRL12,Russomanno_JSTAT13,Russomanno_EPL15,russomanno_JSTAT15,
Lazarides_PRL14,Lazarides_PRE14,Batisdas_PRA12,mori_15,mori_AnnP,Knappa_preprint15,Lindner_Nat11,LS_15, schn_MBL_per,sciarmata_EPL14,Buco_arXiv15,angelo_arXiv16}, as well as proposals for nonequilibrium quantum phase transitions \cite{Gemelke_PRL05,Lignier_PRL07,Zenesini_PRL09,Eckardt_PRL051} with special interest on the preparation
of Floquet time-crystals (referred also as $\pi$-spin glasses)~\cite{Nayak_PRL16,zhang_16:preprint,choi_16:preprint,Vedika_PRL16,Vedika_PRB16,time_crystals_exp:preprint,Vedika_la_santa,moessner2017equilibration} and symmetry-protected topological phases~\cite{Oca_PRB09,Erez_PRB10,inoue2010photoinduced,lindner2011floquet,Jotzu_Nat14,Iacopo_preprint15,Emanuele_arXiv15,d2015dynamical,Toni_arXiv16,
seetharam2015controlled,Liang_PRL06,Manisha_PRB13,Vedika_PRL16,norman_16:preprint}. In particular, Floquet topological insulators have been recently experimentally realized in optical waveguides~\cite{rechtsman2013photonic} and acoustic crystals\cite{fleury2015floquet}. These phases are analogous to the topological band insulators of non-interacting fermions, whose complete equilibrium classification is given by a known mathematical theory, the K-theory (see Ref.~\ocite{Bernevig:book} for an introduction). 
%
%If interactions are present, the fate of this classification is still unknown. For example, Ref.~\ocite{fidkowski2010effects} presented a model of interacting fermions in which the number of topological phases is reduced from infinity to 8. Employing the theory of matrix-product-states (MPS) Refs.~ \ocite{pollman2012symmetry,pollman2010entanglement,turner2011topological} were able to account for this result a classify all possible topological phases of bosons and fermions in one dimension (see also Chen \etal~\cite{chen2011classification} for an alternative derivation). [Removed because not relevant to the present discusssion]

By definition, topological phases do not possess any local order parameter. Nevertheless, in one-dimension these phases can sometimes be characterized by nonlocal string orders\cite{note_cir}. Nonlocal orders account for correlations between spins that are not detected by a simple two-point correlation function, and reveal the presence of a hidden order parameter. In particular, string orders were discussed in the context of the Haldane gapped phase of integer-spin chains~\cite{Hald1,Hald2,Nis,torre2006hidden}, but are in fact ubiquitous in one-dimensional insulators. They appear for example in spin-$1/2$ chain with next-nearest-neighbor interactions\cite{smacchia_ar11:preprint,degottardi2013majorana} and even in the trivial phase of the one-dimensional Bose-Hubbard model \cite{berg2008rise,endres2011observation}. 

The interest in the behavior of the string order in nonequilibrium situations was risen by Ref.~\ocite{Strinati_PRB16}, where it was shown  that sudden external perturbations can destroy the string order of spin-1 chains. In the context of Floquet systems, Ref.~\ocite{Toni_arXiv16} considered a time-periodic Hamiltonian whose high-frequency expansion corresponds to the cluster-Ising model of Ref.~\ocite{smacchia_ar11:preprint} and displays a string order. In the same spirit, by constructing an effective Hamiltonian in the high-frequency limit, Ref.~\ocite{serena_arx} found a string order in a Fermi-Hubbard model with long-range interactions undergoing a periodic driving.
Furthermore, Ref.~\ocite{Vedika_PRL16} considered disordered systems in the many-body localized phase and found that all the eigenstates undergo a transition from a paramagnetic phase to a spin-glass phase with a nonvanishing string-order parameter.

\begin{figure*}

\definecolor{1}{HTML}{8DD3C7}
\definecolor{2}{HTML}{FFFFB3}
\definecolor{3}{HTML}{BEBADA}
\definecolor{4}{HTML}{FB8072}
\definecolor{5}{HTML}{80B1D3}
\definecolor{6}{HTML}{FDB462}
\definecolor{7}{HTML}{B3DE69}
\definecolor{8}{HTML}{FCCDE5}

\begin{minipage}{0.68\textwidth}
\includegraphics[scale=0.5]{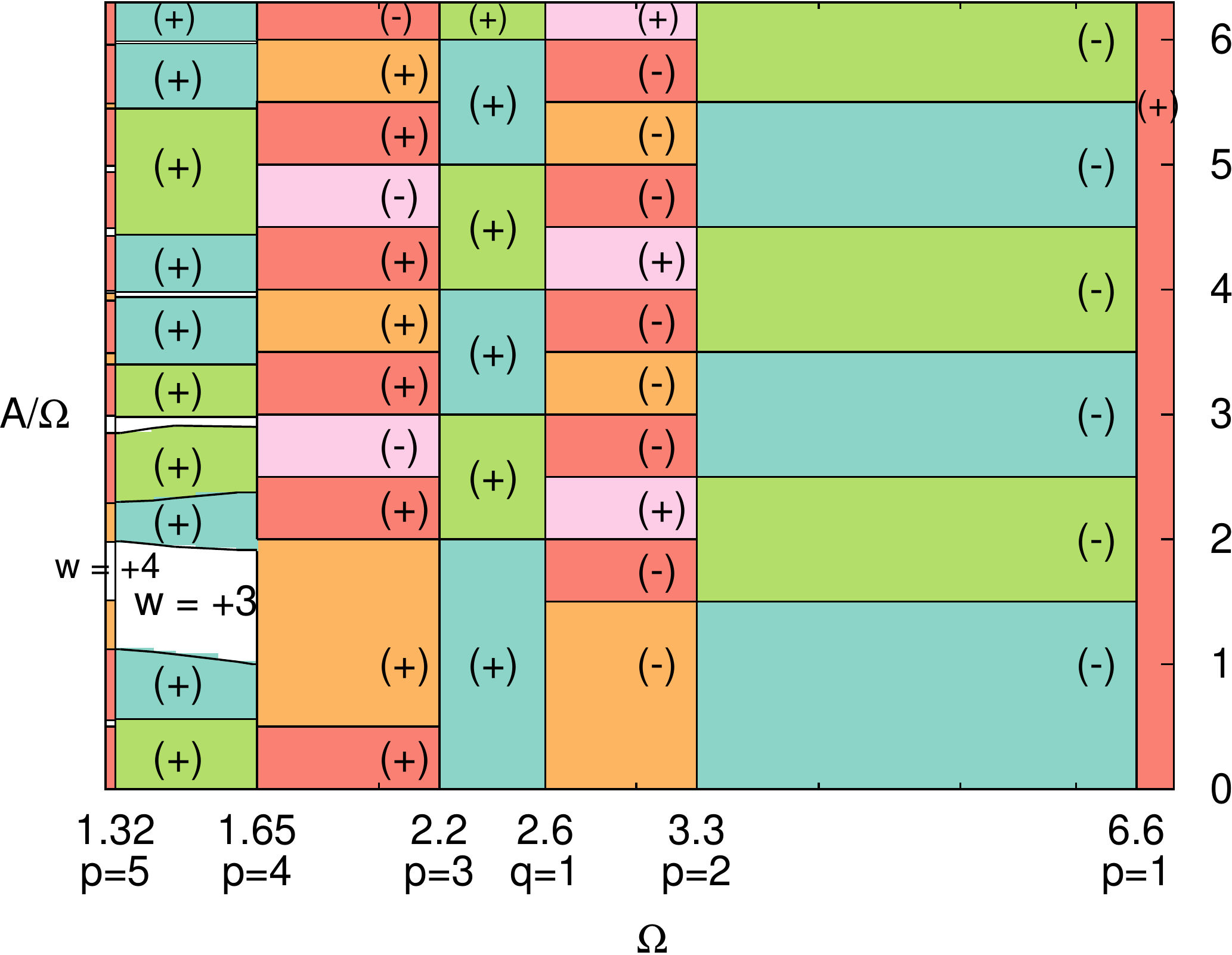}
\end{minipage}
\begin{minipage}{0.3\textwidth}
 \begin{tabular}{|p{1.5cm}|p{1.5cm}|p{1.5cm}|}
  \hline 
Winding number $w$	& Order ~~~~~ parameter 	&  \\
  \hline
$\geq 3$    	&  						& \\
  \hline
2 			&$O^y$              	& \cellcolor{6}\\
  \hline
1			&$S_z$ or $S_z^{\,(-)}$	& \cellcolor{1}\\
  \hline
0			&$O^z$      	   		&  \cellcolor{4}\\
  \hline
-1			&$S_y$ or $S_y^{\,(-)}$ & \cellcolor{7}\\
  \hline
-2			&$O^\beta$  			& \cellcolor{8}\\
  \hline
 \end{tabular} 
\end{minipage}

\caption{(Left panel) Phase diagram of the periodically driven Ising model (\ref{h11}) with $h_0=2.3$. The order parameters and winding numbers were computed at the time-reversal invariant point $t_b=3\tau/4$. The $(+)$ or $(-)$ signs specify if the phase has normal or staggered order parameter. (Right panel) Correspondence among the order parameters at $t_b$ and the values of the winding number $w$.}
\label{diag_fase:fig}
\end{figure*}

The goal of the present manuscript is to study in detail the phase diagram of the periodically driven quantum Ising model, in the absence of disorder. Thanks to the integrability of the model, we are not restricted to the high-frequency limit and we can furthermore extrapolate our analysis to long times and large systems. Following Ref.~\ocite{Emanuele_arXiv15}, we focus on one specific eigenstate of the periodically driven dynamics, the  Floquet Ground State (FGS). %This eigenstate can be obtained by adiabatically reducing the frequency of the driving field, or  turning on its amplitude. 
It shares many properties with the ground state of the static systems. In particular, it has been shown that in integrable models, the entanglement entropy of the FGS generically follows an area law and is bounded in one dimension\cite{Emanuele_arXiv15}. Furthermore, the FGS can undergo sharp phase transitions associated with degeneracies (resonances) of the Floquet spectrum. We will better discuss the properties of this state in the next sections.

We consider the specific case of the clean Ising chain with a time-periodic transverse magnetic field. Previous studies focused on the ground state of effective Hamiltonians
in the limits of high frequency or large amplitude~\cite{Batisdas_PRA12,Toni_arXiv16}, and on the properties of the single quasi-particle states in the fermionic equivalent model\cite{Manisha_PRB13} (defined through the Jordan-Wigner mapping). These studies demonstrated that this system is characterized by a rich phase diagram, including both topologically trivial and nontrivial phases. In Ref.~\ocite{Emanuele_arXiv15} we considered its FGS and showed that this is the specific state undergoing quantum phase transitions~\cite{Note0}. 
Here we significantly extend these results by deriving the full phase diagram of the model. Ref.~\ocite{Emanuele_arXiv15} dealt with the model in the fermionic representation, where the phases have topological nature; here we focus on the model in the spin representation: we are able to show that some of these phases are characterized by a local order parameter, and others by a nonlocal string order parameter (see Fig.~\ref{diag_fase:fig}). 
We observe that the order parameters tend to zero at the transition points, indicating that the transitions are continuous (second order). 
For a system with open boundary conditions the FGS of the topological phases is not unique: in the fermionic representation there are topologically-protected edge states; they are the analog of the Majorana edge states of topological superconductors in one dimension\cite{kitaev2001unpaired,Alicea_RPP12}. \textcolor{black}{As we will show, the transition between any two topological phases can be alternatively described by the vanishing of a local or nonlocal order parameter, a change in the number of Majorana modes, or an integer jump of a topological number. In some of the phases that we find, the FGS breaks time-translation symmetry, like all (or at least an extensive number of) the Floquet states do in the case of time-crystals~\cite{Nayak_PRL16}.}

The paper is organized as follows. In Sec.~\ref{model:sec} we introduce the periodically driven spin-chain model and we briefly discuss how to integrate
it by means of the Jordan-Wigner mapping to a free-fermion model. We then describe its dynamics in terms of the Floquet theory and we precisely define the concept of
Floquet ground state. For completeness of presentation, we give also the details on the construction of the Floquet ground state in general. In Sec.~\ref{topo:sec} we introduce the topological parameters marking our nonequilibrium phases in the fermionic representation
while in Sec.~\ref{parametra:sec} we discuss the order parameters characterizing these transitions in the spin representation. In Sec.~\ref{orpam:sec} we discuss the phase diagram of Fig.~\ref{diag_fase:fig} considering more in detail the properties of the phases and the phase transitions. We find that our system
has a countable infinity of different phases characterized by a pair bulk topological integer numbers: this is consistent with the fact that, thanks to its symmetries, it falls in the \textcolor{black}{BDI} symmetry class with $d=1$ of the classification presented in Ref.~\ocite{Roy_arXiv16} and its \textcolor{black}{classifying} group is $\mathbb{Z}\times\mathbb{Z}$. The two integers characterizing each phase are the winding number $w$ and the loop topological index $n_L$; in Sec.~\ref{OBC:sec} we show that, in systems with open boundary conditions, some edge states appear and their number is related to $w$ and $n_L$. \textcolor{black}{In this section we discuss also the relation between the existence of Majorana edge modes and the time-translation symmetry breaking displayed by the FGS in some of the phases of the system.}

In Sec.~\ref{ramping_correlation:sec} we consider how to prepare these nontrivial phases. %This is impossible to do it directly; o
One
possibility is to start from a vanishing amplitude (or an infinite frequency) and adiabatically change the parameters of the driving until reaching the transition to the considered nontrivial phase (see Ref.~\ocite{Baninno_arXiv16} for a discussion of the same protocol in driven non-integrable systems). Unfortunately, each transition corresponds to
a vanishing gap of the Floquet Hamiltonian: the adiabatic theorem fails and we always reach a state without long-range order. Nevertheless, we find that the
finite correlation length along which the order extends after the ramping protocol 
 scales with the duration of the protocol in a way similar to the standard Kibble-Zurek
effect. %We have order not on an infinite range, but on a range as long as we like. 
It is therefore possible to reach any correlation length by choosing a sufficiently slow crossing rate.
In the Sec.~\ref{concludo:sec} we draw our conclusions. The three appendices discuss the symmetries of the FGS (\ref{time-rev:app}), offer an approximate analytical formula for the positions of the FGS transitions (\ref{resonances:sec}), and explain how to numerically evaluate the Floquet Hamiltonian in absence of translation invariance (\ref{Bogoliubov:sec}).
\section{The spin model and its fermionic description} \label{model:sec}
%-----------------------------------------------------------------------------------------------------------
%
In the present work we consider the dynamics of a uniform quantum Ising chain in the presence of a time-periodic transverse field, described by the Hamiltonian
%After introducing the model, we discuss first the LRT approximation, then the exact Floquet analysis, and finally we compare the two results.
%
%
%
%\begin{equation}  \label{h11}
%  \hat{H}(t)=-\frac{J}{2}\sum_{j=1}^{L}\left(\sigma_j^z\sigma_{j+1}^z+h_j(t)\sigma_j^x\right) \;.
%\end{equation}
%
\begin{equation}  \label{h11}
  \hat{H}(t) =\frac{1}{2}\sum_{j=1}^{L}\left(-J\hat{\sigma}_j^z\hat{\sigma}_{j+1}^z + h(t) \hat{\sigma}_j^x\right)  \;.
\end{equation}
Here $\hat{\sigma}^{x,z}_j$ are spins (Pauli matrices) at site $j$ of a chain of length $L$ with boundary conditions which can be periodic 
(PBC) $\hat{\sigma}^{x,z}_{L+1}=\hat{\sigma}^{x,z}_1$ or open (OBC) $\hat{\sigma}^{x,z}_{L+1}=0$, and $J$ is a longitudinal coupling ($J=1$ in the following). The transverse field is taken uniform and time-periodic, $h(t)=h(t+\tau)$. %For concreteness, we consider a step-wise driving, with a magnetic field jumping between $h_0+A$ and $h_0-A$. Furthermore, for concreteness we arbitrarily set $h_0=2.3$. Thanks to the analytical insights provided in this paper, our analysis can be straightforwardly extended to any periodic drive.
Although in our numerical calculations we will focus on a specific driving protocol~\cite{Noteb}, the analytical insights provided in this paper allow to straightforwardly extend
our analysis to any periodic drive.

%and $h_j(t)$ is a transverse field. %(in the following we set $J=\hbar=1$). 
%

Let us now review the equilibrium properties of the Hamiltonian (\ref{h11}) with a static and homogeneous transverse field, $h(t)=h_0$. This model has two gapped phases:
a ferromagnet ($\left|h_0\right|<1$) and a paramagnet ($\left|h_0\right|>1$), separated by a quantum phase transition at $h_c=1$. This result is not difficult to find because the Hamiltonian \eqref{h11} can be transformed, through a Jordan-Wigner transformation (see Refs.~\ocite{Lieb_AP61,Pfeuty_AP70}), 
into a quadratic fermionic form. %, which transforms  $\hat{H}(t)$ to a . 
In the case of PBC {for the spins}, we can quite simplify the analysis: going to $k$-space,
$\hat{H}(t)$ becomes a sum of two-level systems
\begin{eqnarray} \label{Ht:eqn}
  \hat{H}(t) &=&  \sum_{k>0}^{\rm ABC}
                    \left(\begin{array}{cc}
			\opcdag{k} & \opc{-k}
		\end{array}\right)
	  \mathbb{H}_k(t)
	\left(\begin{array}{c}
			\opc{k} \\
			\\
			\opcdag{-k}
		\end{array}\right) \nonumber\\
\hspace{2mm} &\mbox{with}& \hspace{4mm} 
\mathbb{H}_k(t)\equiv \left(\begin{array}{cc}
			\epsilon_k(t) &-i\Delta_k\\
			i\Delta_k^*&-\epsilon_k(t)
		\end{array} \right)
%      \sum_{k}^{\rm ABC} \big[ \epsilon_k(t)  \left(c_k^\dagger c_k-c_{-k} c_{-k}^\dagger\right) 
%                            - i\Delta_k \left(c_k^\dagger c_{-k}^\dagger-c_{-k} c_k\right)\big] \;,
\end{eqnarray}
where $\epsilon_k(t) = h(t)-\cos{k}$, $\Delta_k=\sin{k}$, and the sum over $k$ is restricted to positive $k$'s of the form 
$k=(2n+1)\pi/L$ with $n=0,\ldots,L/2-1$, {and $L$ even}, corresponding to anti-periodic boundary conditions 
(ABC) for the fermions~\cite{Lieb_AP61,Note2}.
In the following we will refer to such a set of $k$ as $k\in {\rm ABC}$~\cite{Note2}.
The Hamiltonian can be block-diagonalized in each $k$ sector with instantaneous eigenvalues
%Each $\hat{H}_k(t)$ acts on a 2-dim Hilbert space generated by $\{ c_k^\dagger c_{-k}^\dagger \ket{0}, \ket{0} \}$,
%and can be represented in that basis by a $2\times 2$ matrix
%$H_k(t)=\epsilon_k(t) \sigma^z + \Delta_k \sigma^y$, with instantaneous eigenvalues 
%
%\begin{equation} \label{autoval:eqn}
$\pm E_k(t)=\pm\sqrt{\epsilon_k^2(t)+\Delta_k^2}$;
%\end{equation}
%
%In the same representation, the unperturbed (critical) Hamiltonian is given by
%%
%\begin{equation} \label{H0:eqn}
%\hat{H}_0 = \sum_k^{\textrm{ABC}} \hat{H}_k^0 = \sum_k^{\textrm{ABC}} 
%                  \left(\begin{array}{cc}
%			\opcdag{k} & \opc{-k}
%		\end{array}\right)
%	\left(\begin{array}{cc}
%			1-\cos(k) & -i\sin(k)\\
%			\\
%			i\sin(k) & \cos(k)-1
%		\end{array}\right)
%	\left(\begin{array}{c}
%			\opc{k} \\
%			\\
%			\opcdag{-k}
%		\end{array}\right)\;,	%\nonumber
%\end{equation}
%%
%with eigenvalues given by  $\pm\epsilon_k^0 = \pm 2\sin(k/2)$.
%This immediately implies that the natural resonance frequencies are at $\pm 2\epsilon_k^0$, which in our units are between $-4$ and $4$.
%
%We assume that at time $t=0$ the coherent evolution starts with the system in 
the ground state of the Hamiltonian $\hat{H}$ with a fixed value of the field $h_0$ has a BCS-like form
%This ground state 
%
\begin{equation} \label{ground}
\ket{\psi_{\rm GS}} = \prod_{k>0}^{\rm ABC} \ket{\psi_k^0} = \prod_{k>0}^{\rm ABC} \left(v_k^{0}+u_k^{0}\opcdag{k} \opcdag{-k}\right) \ket{0}\,,
\end{equation}
with $v_k^{0}=\cos(\theta_k/2)$ and $u_k^{0}=i\sin(\theta_k/2)$ expressed in terms of an angle $\theta_k$ defined by $\tan{\theta_k} = (\sin{k})/(h_0-\cos{k})$.
%
%We describe here the exact 
%\textcolor{black}{The Hamiltonian Eq.~\eqref{h11} is invariant under the group of transformations}
%
%\begin{equation}
%  \hat{V}(n)=\exp\left(-in\frac{\pi}{2}\sum_j\hat{\sigma}_j^x\right)\quad{\rm with}\quad n\in\mathbb{Z}\,.
%\end{equation}
%
%\textcolor{black}{This is the group of the rotations of an integer number of $\pi$ around the $x$-axis and it can be identified with the group $\mathbb{Z}$ of
%the integer numbers: this Hamiltonian is said to belong to the $\mathbb{Z}$-symmetry class~\cite{Kane_RMP}.} %{\bf Ho cancellato queste due frasi perche' non le capisco: innanzitutto $V(4)==1$, quindi l'identificazione di $V(n)$ con il gruppo $\mathbb{Z}$ non e' univoca. In secondo luogo il lavoro che citi parla di modelli fermionici e in particolare non specifica la simmetria del modello (1). In ultimo, non capisco che relazione ci sia tra la prima parte della frase e la seconda. Ne parliamo via Skype?} 
%\textcolor{black}{Fixing the time and seeing the instantaneous Hamiltonian~\eqref{Ht:eqn} as a time-independent object -- as appropriate for the evaluation of the eigenstates}, 

In the equilibrium case, moreover, %($h(t)=h_0$)  
the Hamiltonian~\eqref{Ht:eqn} is invariant under both particle-hole and time-reversal symmetries. The former symmetry is valid for any Hermitian operator that can be written in the form \eqref{Ht:eqn}, with a generic $\mathbb{H}_k$. Mathematically, this symmetry can be defined as
\begin{equation} \label{tras1:eqn}
  \tau_x\mathbb{H}^*_k\tau_x = - \mathbb{H}_{-k}\;,
\end{equation}
where $\tau_x$ flips particles and holes and acts as a Pauli matrix in Nambu space. This relation can be shown to apply to any BCS-like Hamiltonian, using the anti-commutation relations for fermions.
Time reversal symmetry is equivalent to
\begin{equation} \label{tras2:eqn}
  \mathbb{H}_k = \mathbb{H}_{-k}^*\,.
\end{equation}
and applies to the present case as long as $\Delta_k=-\Delta_{-k}$.
These two conditions imply that the coefficients $u_k^0$ and $v_k^0$ of the ground state enjoy the property
\begin{equation} \label{timer:eqn}
  \Real\left[u_k^0(v_k^0)^*\right]=0\,.
\end{equation}
%{\bf Question: can we show the opposite? If Eq. (7) is satisfied can we deduce (5) and (6)?} 
%
According to the K theory, static one dimensional systems with time-reversal and particle-hole symmetries can at most support a numerable infinity of distinct topological phases (they are said to belong to the $\mathbb{Z}$
symmetry class). As we will see below, this property can still be valid also in the time-periodic case.

%
%In the notation of Sec.~\ref{linear-response:sec}, $\hat{H}(t)=\hat{H}_0 + v(t) \hat{A}$ where $\hat{H}_0$ is the homogeneous model with transverse field $h$ 
%(which we will set for convenience to critical value $h=h_c=1$) and $\hat{A}=\hat{M}_l=\sum_{j=1}^l \sigma_j^x$ is the transverse magnetization of a region comprising $l$ sites.
%
%We start discussing the extensive case with $l=L$ (the periodic driving acts on the whole chain), where translational invariance
%simplifies the analysis considerably, since $\hat{A}=\hat{M}_L=\sum_{j=1}^L \sigma_j^x$. 
%Further technical details for the general nontranslationally invariant case are contained in \ref{Bogoliubov:sec}.
%
%If we assume 

Let us move now to the periodically driven case. To analyze the system in this case, it is convenient to perform a Floquet analysis, which we are going to elucidate in its fundamental lines. (See Refs.~\ocite{Hanggi_book,Shirley_PR65, Hausinger_PRA10} for an introduction to Floquet theory and Refs.~\ocite{Russomanno_PRL12,Russomanno_JSTAT13} for more details about the present case.)
  %The evolution of the system with a time-periodic $h(t)$ can be naturally described through a Floquet analysis \cite{Russomanno_PRL12,Russomanno_JSTAT13}. 
%,%as an exemplification of the general arguments of Section \ref{floquet:sec}. 
%In~\ref{Bogoliubov:sec} we show how to extend this picture to the case of OBC or inhomogeneous couplings; 
%in this Sec.~we focus on the case of PBC {for the spins} because it is more transparent and instructive.
The state of the system at all times can be written in a BCS form
\begin{equation} \label{state:eqn}
  \ket{\psi(t)}=\prod_{k>0}^{\rm ABC} \ket{\psi_k(t)}=\prod_{k>0}^{\rm ABC} \left(v_k(t)+u_k(t)\opcdag{k} \opcdag{-k} \right) \ket{0} \;,
\end{equation}
where the functions $u_k(t)$ and $v_k(t)$ obey the Bogoliubov-de Gennes equations 
\begin{equation} \label{deGennes:eqn}
    i\hbar \frac{d}{dt}\left(\begin{array}{cc}
			u_k(t)\\v_{k}(t)
           		\end{array} \right)
%		= H_k(t) \left(\begin{array}{cc}
%		          	u_k(t)\\v_{k}(t)
%		               \end{array} \right)
		= \mathbb{H}_k(t)
		\left(\begin{array}{cc}
			u_k(t)\\v_{k}(t)
		\end{array} \right) 
	\;,
\end{equation}
%
%with initial values $v_k(0)=v_k^{0}$ and $u_k(0)=u_k^{0}$.
%, because at time $t=0$ the system is in the ground state. 
%The dynamics is quite clearly factorized in the two-dimensional subspaces generated by $\{ \opcdag{k} \opcdag{-k} \ket{0}, \ket{0} \}$.
%
%The transverse magnetization operator $\hat{M}_L$ reads, in terms of Jordan-Wigner fermions, as 
%$\hat{M}_L=\sum_{k>0}^{\textrm{ABC}} \hat{m}_k$ where $\hat{m}_k = 2 \left( c_{-k} c_{-k}^\dagger - c_k^\dagger c_k \right)$.
%Using this, we can express the average transverse magnetization density at time $t$, in the thermodynamic limit, as:
%
%\begin{equation} \label{magnets}
%  m(t) = \int_0^\pi \! \frac{\ud k}{2\pi} \; \bra{\psi_k(t)} \hat{m}_k \ket{\psi_k(t)} \;.
%\end{equation}
%
%The previous analysis applies to a general $h(t)$. 
with $\mathbb{H}_k(t)$ defined in Eq.~\eqref{Ht:eqn}. For a time-periodic $h(t)=h(t+\tau)$, we can find a basis of states which are $\tau$-periodic 
``up to a phase'' in each $k$-subspace. 
These are the Floquet states $\ket{\psi_k^{\pm}(t)}=\nep^{\mp i \mu_k t}\ket{\phi_k^{\pm}(t)}$: the $\ket{\phi_k^{\pm}(t)}$ 
are $\tau-$periodic, and are called Floquet modes, while the $\pm\mu_k$ are real and are called quasi-energies.
The two quasi-energies 
have opposite signs because $\mathbb{H}_k(t)$ has vanishing trace~\cite{Russomanno_PRL12,Russomanno_JSTAT13}
and are defined up to translations of multiples of the driving frequency $\Omega=2\pi/\tau$, in complete analogy with Bloch quasi-momenta in periodic solids:
each interval of amplitude $\Omega$ is a ``Brillouin zone'' for the quasi-energies. Details on how to compute Floquet modes and quasi-energies in this case are given in \cite{Russomanno_PRL12,Russomanno_JSTAT13} and the related supplementary material. 

The stroboscopic dynamics of the system (at times $t_n=\delta t+n\tau$,  with $n\in \mathbb{Z}$) is induced by an effective 
Hamiltonian $\hat{H}_{F}(\delta t)$ called Floquet Hamiltonian 
\begin{equation}
  \hat{U}(\delta t+n\tau,\delta t)=\nep^{-i\hat{H}_{F}(\delta t)n\tau}\,.
\end{equation}
Because the problem can be factorized in $2\times2$ subspaces, we can write the Floquet Hamiltonian for any time $\delta t$ as
\begin{align} \label{HF:eqn}
  \hat{H}_{F}(\delta t) &=  \sum_{k>0}^{\rm ABC}
                    \left(\begin{array}{cc}
			\opcdag{k} & \opc{-k}
		\end{array}\right)
	  \mathbb{H}_{k\,F}(\delta t)
	\left(\begin{array}{c}
			\opc{k} \\
			\\
			\opcdag{-k}
		\end{array}\right)\quad \mbox{with} \nonumber\\
 %\hspace{2mm}  
\left(\mathbb{H}_{k\,F}(\delta t)\right)_{i,j}&\equiv \sum_{\alpha=\pm}\mu_k^\alpha 
   \left\langle\chi_i^{(k)}\right.\ket{\phi_k^{\alpha}(\delta t)}
     \bra{\phi_k^{\alpha}(\delta t)}\left.\chi_j^{(k)}\right\rangle\,,
%      \sum_{k}^{\rm ABC} \big[ \epsilon_k(t)  \left(c_k^\dagger c_k-c_{-k} c_{-k}^\dagger\right) 
%                            - i\Delta_k \left(c_k^\dagger c_{-k}^\dagger-c_{-k} c_k\right)\big] \;,
\end{align}
%
%The state of the system at any time $t$ can then be expanded as
%
%\begin{equation} \label{expansion:eqn}
%  \ket{\psi_k(t)} = r_k^+ \nep^{-i\mu_k t} \ket{\phi_k^+(t)} + r_k^- \nep^{i\mu_k t} \ket{\phi_k^-(t)} \;,
%\end{equation}
%
%where $r_k^{\pm} = \left\langle \phi_k^{\pm}(0) \right. \ket{\psi_k(0)}$ are the overlap factors between the 
%initial state $\ket{\psi_k(0)}$ and the Floquet modes $\ket{\phi_k^{\pm}(t)}$. 
%
where $i,j=1,2$ and we have defined $\ket{\chi_1^{(k)}}\equiv \opcdag{k} \opcdag{-k} \ket{0}_k$ and $\ket{\chi_2^{(k)}}\equiv \ket{0}_k$.
Note that the Floquet Hamiltonian is by construction particle/hole symmetric; its behavior under time-reversal will be explored below (see Appendix~\ref{time-rev:app}
for details on the symmetry properties of this object). 

In this paper we consider a specific eigenstate of the Floquet Hamiltonian, the Floquet ground state (FGS), as defined in Ref.~\ocite{Emanuele_arXiv15}. This state is the adiabatic continuation to finite frequency of the ground state at infinite frequency: in this limit the Floquet Hamiltonian simply coincides with the time-averaged Hamiltonian. For this specific system, we find that the FGS coincides with the ground state of the Floquet Hamiltonian as defined in Eq.~\eqref{HF:eqn}. If we consider the quasi-energies in the first Brillouin zone $[-\Omega/2,\Omega/2]$, we
can see that the spectrum of the Floquet Hamiltonian $\hat{H}_{F}(\delta t)$ is independent of $\delta t$ and is made by two bands. The FGS is obtained by completely filling the lower band~\cite{Note_ex}
 %\footnote{By the way what is the lower and what is the upper band is not univocally defined: it depends on the choice of the Brillouin zone. With a different choice, the ground state of the Floquet Hamiltonian would be {$\prod_{k>0}^{\rm ABC} \ket{\phi_k^{-}(0)}$}. This state shows exactly the same phase transitions of the Floquet Ground state (and is the adiabatic continuation of the mostly excited state of the static model), so this point is not very crucial. }.
and can be explicitly written as
\begin{equation} \label{state:eqn}
  \ket{\Psi_{\rm FGS}(t)}=\nep^{-i\overline{\mu}_{\rm FGS}t}\prod_{k>0}^{\rm ABC} \left(v_{k\,P}^-(t)+u_{k\,P}^-(t)\opcdag{k} \opcdag{-k} \right) \ket{0}_k \,,
\end{equation}
where $\overline{\mu}_{\rm FGS}=-\sum_{k>0}^{\rm ABC}\mu_k$ is the corresponding many-body quasi-energy and $v_{k\,P}^-(t)$, $u_{k\,P}^-(t)$ are the time-periodic
amplitudes of the ground state of $\mathbb{H}_{k\,F}(t)$ in the basis $\{\ket{0}_k,\opcdag{k} \opcdag{-k} \ket{0}_k\}$. To be more precise, we should say that
our analysis focuses on the Floquet ground mode: this is
the $\tau$-periodic state coinciding with the Floquet ground state up to a phase
\begin{equation} \label{statem:eqn}
  \ket{\Phi_{\rm FGS}(\delta t)}=\prod_{k>0}^{\rm ABC} \left(v_{k\,P}^-(\delta t)+u_{k\,P}^-(\delta t)\opcdag{k} \opcdag{-k} \right) \ket{0}_k \,.
\end{equation}
Being this state periodic, we can restrict $\delta t \in[0,\tau]$. Nevertheless, the phase factor will always simplify, so we will indifferently talk about Floquet
ground state or Floquet ground mode. Moreover, because in this model the FGS is the ground state of the Floquet Hamiltonian, we will indifferently say that the phase transitions are of the Floquet Hamiltonian or of the Floquet ground state/mode.

\begin{figure}
\centering
\includegraphics[scale=0.8]{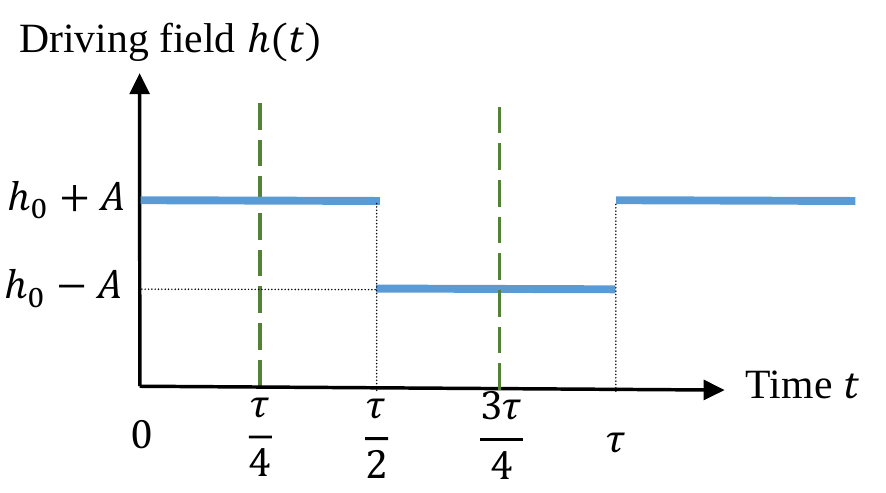}
\caption{The driving protocol considered in the present study, with period $\tau=2\pi/\Omega$. The dashed vertical lines indicate the two time-reversal symmetric points at $t_a=\tau/4$ and $t_b=3\tau/4$.}
\label{fig:driving}
\end{figure}

We conclude this section reporting, for completeness of presentation, the details on the construction of the Floquet ground state in general,
already discussed in Ref.~\ocite{Emanuele_arXiv15}. The Floquet ground state is defined as the adiabatic continuation at finite frequency of the ground state of the high-frequency 
time-averaged Hamiltonian. %; we explicitly discuss the prescription to construct this continuation. 
In the high frequency limit, the ground state of the time-averaged Hamiltonian is a legitimate Floquet state~\cite{Emanuele_arXiv15,Russomanno_PRL12}.
We can define the Floquet state which adiabatically continues it to lower frequencies with a prescription which goes as follows.

%We explicitly emphasize that, in general, the FGS is not the ground state of any Hamiltonian and inherits the word ``ground'' from the ground
%state which adiabatically continues. We agree with the Referee that this may seem a slight abuse of notation, but the FGS shares important 
%properties with the ground state of a static Hamiltonian (second order phase transitions, non-extensive entropy) and then we have decided to
%use this name. We thank again the referee: she/he has given us the possibility to make clear this point which in the old version of the manuscript
%was not discussed with the due attention.
%We explain in detail how to do the adiabatic continuation from
%the high frequency limit to a finite frequency. 
%
Let us start considering the adiabatic continuation of the FGS for an infinitesimal change of frequency.
Our first step is fixing the phase of the periodic motion $\theta\equiv t/\tau\in[0,2\pi]$ and
 considering a Floquet mode
at a frequency $\Omega$: $\ket{\Phi_{\Omega,\,\gamma}(\tau\theta)}$. If we take a frequency infinitesimally
smaller $\Omega-\delta\Omega$, the adiabatic continuation is the Floquet mode at the new frequency 
$\ket{\Phi_{\eta,\Omega-\delta\Omega}\left(\tau\left(1+\frac{\delta\Omega}{\Omega}\right)\theta\right)}$ chosen so that the overlap
\begin{displaymath}
  \left|\left\langle\Psi_{\eta,\Omega-\delta\Omega}\left(\tau\left(1+\frac{\delta\Omega}{\Omega}\right)\theta\right)\right.\ket{\Psi_{\gamma,\Omega}(\tau\theta)}\right|^2
\end{displaymath}
is {\em maximum}. ``Integrating'' this infinitesimal step from the ground state of the time-averaged Hamiltonian
in the high frequency limit %$\Omega=\infty$
to the desired value of $\Omega$, and repeating the procedure for all the values
of $\theta\in[0,2\pi]$, we obtain the full time-dependence of the desired Floquet ground mode (the Floquet ground state can be obtained
by multiplying the phase factor $\nep^{-i\overline{\mu}_{\rm FGS} t}$, where $\overline{\mu}_{\rm FGS}$ is the corresponding quasi-energy). %ractically, we do this operation numerically-
%starting from a very high value of $\Omega$).
%The Floquet modes can be constant as a degenerate case, but in general they are time-periodic: % and then
%also the adiabatic continuation is in general time dependent.

%\textcolor{black}{ %There are many possible ways to do the adiabatic continuation until the final value of $\Omega$: we can choose another path in which we vary also other parameters (like $A$ and $h_0$ in our case). We can see that the resulting the Floquet ground state is independent of the path taken, because the adiabatic continuation is a geodesic curve in a sub-manifold of the Hilbert space with respect to the metric induced by the real part of the quantum geometric tensor~\cite{Zanardi_PRL99,Ammammata_PRE11}. 
%The way the Floquet ground state is constructed works also if we have to cross a resonance, because it is a process which
%does not involve any dynamics, but only the knowledge of the Floquet states at any frequency (which can be obtained, admittedly,
%only for integrable systems like the one we are studying.}

Applying the procedure we have described to this model (see Eq.~\eqref{h11}), we have numerically
verified that we obtain the state in Eq.~\eqref{statem:eqn}~\cite{Emanuele_arXiv15}. We have done it with a finite system: if we take a longer chain we
need a smaller step $\delta\Omega$ to go across the quantum phase transition points. Nevertheless, going to the limit $L\to\infty$ we get a perfectly defined state, and the result does not change if we make the adiabatic continuation along a different path (one in which we change also $A$, for instance).
In the next Section we discuss the details of the topological parameters characterizing the Floquet ground state in our model.

%{\bf Emanuele: Ho riportato in questa sezione la discussione sugli ordini topologici fermionici . La logia e' sezione II = fermioni e sezione III = spin.}
\section{Topological parameters in the fermionic representation} ~\label{topo:sec}
%\prod_{k>0}^{\rm ABC} \ket{\phi_k^{-}(t)}=
For the topological classification of Floquet quantum phases it is important to distinguish driving protocols with at least one time-reversal invariant point $t_r$ per period~\cite{Note4}, from those that are not time-reversal invariant \cite{Manisha_PRB13}. In this work we specifically consider the former case: our numerical calculations refer to the protocol depicted in Fig.~\ref{fig:driving}, which has two time-reversal invariant points. At these points the Floquet Hamiltonian obeys the time reversal symmetry and the particle-hole symmetry in a form identical to Eqs.~\eqref{tras1:eqn} and \eqref{tras2:eqn} (see Appendix~\ref{time-rev:app} for details) and then $u_{k\,P}^-(t_r)$ and $v_{k\,P}^-(t_r)$ satisfy Eq.~(\ref{timer:eqn}) 
\begin{equation} \label{trasimm:eqn}
  \Real\left(u_{k\,P}^-\left(t_r\right)\left(v_{k\,P}^-\left(t_r\right)\right)^*\right)= 0\,,
\end{equation}
and the FGS is time-reversal symmetric. According to the K theory, the combination of particle/hole and time-reversal symmetries lead to an infinite number of distinct topological phases. These phases are characterized by the winding number $w$~\cite{Emanuele_arXiv15}, which has a simple geometrical interpretation: Thanks to Eq.~\eqref{trasimm:eqn}, the Bloch vector representative
of the spinor $\left(\begin{array}{c}u_{k\,P}^-(t_r)\\v_{k\,P}^-(t_r)\end{array}\right)$ lays on the $xy$ plane. The winding number $w$ simply counts the number of revolutions that this vector
performs around the origin, as $k$ varies between 0 and $2\pi$. This quantity is equivalently given by the Berry phase acquired by the spinor during this revolution~\cite{Berry}, divided by $\pi$. 

In contrast, at times where the system is not time-reversal invariant, Eq.~\eqref{tras2:eqn} is not satisfied, and the Floquet Hamiltonian is only characterized by the particle-hole symmetry, Eq.~\eqref{tras1:eqn}. 
In this case, according to the K-theory, the Floquet Hamiltonian belongs to the $\mathbb{Z}_2$ symmetry class: it can support at most two
different topological phases. We will come back to this point towards the end of the next section. These phases are characterized by a $\mathbb{Z}_2$ topological invariant\cite{Alicea_RPP12} $\nu=\pm 1$. From a geometrical perspective, in our model we observe that $\nu=+1$ if both $k=0$ and $k=\pi$ are associated with the same pole of the Bloch vector, and $\nu=-1$ if the are associated with opposite poles. Note that the $\mathbb{Z}$ symmetry class is a subset of the $\mathbb{Z}_2$ symmetry class: when $w$ is defined, its parity corresponds to $\nu$ (i.e. $\nu=1$ for even $w$s and $\nu=-1$ for odd ones). In the periodically driven case, there is also the loop topological index $n_L$~\cite{Roy_arXiv16} which we will better discuss in Section~\ref{OBC:sec}. We will see that the countable infinity of phases we find corresponds to a \textcolor{black}{classifying} group $\mathbb{Z}\times\mathbb{Z}$ and each phase is characterized by the pair of indices $(w,n_L)$.

%The existence of topological phase transitions in the FGS of this model was first noticed in Ref.~\ocite{Emanuele_arXiv15}. There we observed that, whenever the gap in the Floquet spectrum closed, the winding number (to be defined below) jumped by an integer, indicating the transition to a new topological phase. Here we rather focus on the spin model and demonstrate that some phases are actually characterized by a local order parameter, while others are genuinely topological:
%they are charcterized by string order.

%
%................................................................................................................................................................%
%''''''''''''''''''''''''''''''''''''''''''''''''''''''''''''''''''''''''''''''''''''''''''''''''''''''''''''''''''''''''''''''''''''''''''''''''''''''''''''''''%
\section{Local and nonlocal spin order parameters} \label{parametra:sec}
We now move to the characterization of the different phases in terms of spin observables. The simplest observable is $\langle{\hat{\sigma}^x_j\rangle}$, which corresponds to the local density of Jordan-Wigner fermions at site $j$. In contrast, the magnetizations along the $y$ and $z$ directions cannot be directly written in terms of local fermionic objects. Their value can nevertheless be obtained as the infinite-distance limit of long-range correlators (see for instance
~\cite{Patasinski:book}):
\begin{equation} \label{local_par:eqn}
 S_z=\frac{1}{2}\sqrt{\lim_{l\to\infty}\mean{\hat{\sigma}_j^z\hat{\sigma}_{j+l}^z}}\quad S_y=\frac{1}{2}\sqrt{\lim_{l\to\infty}\mean{\hat{\sigma}_j^y\hat{\sigma}_{j+l}^y} }\;.
\end{equation}
A similar analysis can be performed for the anti-ferromagnetic order parameters in the $y$ and $z$ direction:
{\small
\begin{eqnarray} \label{local_stag:eqn}
 S_z^{\,(-)}&=&\frac{1}{2}\sqrt{\lim_{l\to\infty}(-1)^l\mean{\hat{\sigma}_j^z\hat{\sigma}_{j+l}^z}}\nonumber\\
  S_y^{\,(-)}&=&\frac{1}{2}\sqrt{\lim_{l\to\infty}(-1)^l\mean{\hat{\sigma}_j^y\hat{\sigma}_{j+l}^y}}\,.
\end{eqnarray}
}

As we mentioned in the introduction, topological phases in one dimension can be characterized by string orders~\cite{note_cir}. A simple type of string order can be introduced by applying to the system the duality transformation~\cite{smacchia_ar11:preprint}
\begin{eqnarray}
  \hat{\mu}_j^x&=&\hat{\sigma}_j^z\hat{\sigma}_{j+1}^z\nonumber\\
  \hat{\mu}_j^y&=&-\left(\prod_{k=1}^{j-1}\hat{\sigma}_{k}^x\right)\hat{\sigma}_{j}^y\hat{\sigma}_{j+1}^z\nonumber\\
  \hat{\mu}_j^z&=&\prod_{k=1}^j\hat{\sigma}_{k}^x\;.
\end{eqnarray}
This duality transformation maps the Hamiltonian~\eqref{h11} to
\begin{equation}
   \hat{H}(t)=-\frac{h(t)}{2}\sum_j\left[\frac{1}{h(t)}\hat{\mu}_j^x-\hat{\mu}_{j-1}^z\hat{\mu}_{j}^z\right]\,.
\end{equation}
Note that the resulting Hamiltonian has the same form as the original one, with $h(t)\to1/h(t)$. At equilibrium this property can be for instance invoked to explain why the transition between the ferromagnetic and paramagnetic phases occurs precisely at the self-dual point $|h_0|=1$, where $h_0 = 1/h_0$. For a time-dependent problem, the duality allows us to map the periodic modulations of the magnetic field considered in Ref.~\ocite{Emanuele_arXiv15} to the periodic modulations of the interaction energy considered in Ref.~\ocite{Toni_arXiv16}. 

The spin correlators in the $\mu$-representation are mapped to highly nonlocal string operators in the $\sigma$-representation~\cite{smacchia_ar11:preprint}:
these are the string order parameters
% \lim_{l\to\infty}(-1)^{l+1}
\begin{eqnarray} \label{string_params:eqn}
    O^y&=&  \mean{\hat{\mu}_j^y\hat{\mu}_{j+l-1}^y}\nonumber\\
       &=& 
      \mean{\hat{\sigma}_j^z\hat{\sigma}_{j+1}^y \left(\prod_{k=j+2}^{j+l-2}\hat{\sigma}_k^x\right)  \hat{\sigma}_{j+l-1}^y\hat{\sigma}_{j+l}^z}\nonumber\\
    O^{y\,(-)} &=& \lim_{l\to\infty}(-1)^{l+1}\lim_{l\to\infty}\mean{\hat{\mu}_j^y\hat{\mu}_{j+l-1}^y}\nonumber\\
    O^z&=&\lim_{l\to\infty} \mean{\hat{\mu}_j^z\hat{\mu}_{j+l}^z}= \mean{\prod_{k=j+1}^{j+l-1}\hat{\sigma}_k^x}\nonumber\\
    O^{z\,(-)}&=&\lim_{l\to\infty} (-1)^l\mean{\hat{\mu}_j^z\hat{\mu}_{j+l}^z}\,.
\end{eqnarray}

As we will explain below, it is useful to introduce a second duality transformation
\begin{eqnarray}
  \hat{\beta}_j^x&=&\hat{\sigma}_j^y\hat{\sigma}_{j+1}^y\nonumber\\
  \hat{\beta}_j^y&=&\left(\prod_{k=1}^{j-1}\hat{\sigma}_{k}^x\right)\hat{\sigma}_j^z\hat{\sigma}_{j+1}^y\nonumber\\
  \hat{\beta}_j^z&=&\prod_{k=1}^j\hat{\sigma}_{k}^x\;,
\end{eqnarray}
which transforms the Hamiltonian \eqref{h11} to
\begin{equation}
  \hat{H}(t) =\frac{1}{2}\sum_{j=1}^{L}\left(J\hat{\beta}_{j-1}^z\hat{\beta}_j^x\hat{\beta}_{j+1}^z + h(t) \hat{\beta}_{j-1}^z\hat{\beta}_{j}^z\right)  \;.
\end{equation}
After an appropriate Jordan-Wigner transformation followed by a Fourier transform, introducing the $\opb{k}$ fermionic operators, this Hamiltonian acquires the free-fermion BCS form
\begin{eqnarray} \label{Htb:eqn}
  \hat{H}(t) &=&  \sum_{k>0}^{\rm ABC}
                    \left(\begin{array}{cc}
			\opbdag{k} & \opb{-k}
		\end{array}\right)
	  \mathbb{H}_k(t)
	\left(\begin{array}{c}
			\opb{k} \\
			\\
			\opbdag{-k}
		\end{array}\right)\quad \mbox{with} \nonumber\\
%\hspace{2mm} &}& \hspace{4mm} 
\mathbb{H}_k(t)&\equiv& \left(\begin{array}{cc}
			J\cos(2k)-h(t)\cos k &i h(t) \sin k-iJ\sin(2k)\\
			-i h(t)\sin k+iJ\sin(2k)& h(t)\cos k - J\cos(2k)
		\end{array} \right)\,.\nonumber
%      \sum_{k}^{\rm ABC} \big[ \epsilon_k(t)  \left(c_k^\dagger c_k-c_{-k} c_{-k}^\dagger\right) 
%                            - i\Delta_k \left(c_k^\dagger c_{-k}^\dagger-c_{-k} c_k\right)\big] \;,
\end{eqnarray}
By considering the $\hat{\beta}^y$ correlation functions it is possible to define a new string order parameter
\begin{eqnarray} \label{Obeta:eqn}
  O^\beta&=&  \lim_{l\to\infty}\mean{\hat{\beta}_j^y\hat{\beta}_{j+l-1}^y} \nonumber\\
       &=& \lim_{l\to\infty}(-1)^{l+1}
      \mean{\hat{\sigma}_j^y\hat{\sigma}_{j+1}^z \left(\prod_{k=j+2}^{j+l-2}\hat{\sigma}_k^x\right)  \hat{\sigma}_{j+l-1}^z\hat{\sigma}_{j+l}^y}\,.\nonumber\\
  O^{\beta\,(-)}&=&\lim_{l\to\infty}(-1)^{l+1}\mean{\hat{\beta}_j^y\hat{\beta}_{j+l-1}^y}\;.
\end{eqnarray}

In general, these order parameters have a nonlocal representation in the fermionic language. Their expressions, obtained by means of the Jordan-Wigner
transformation and the Wick's theorem, 
%The expressions for these correlators  and in Appendix~\ref{app_pfaff}
are given in Ref.~\ocite{Barouch_PRA71}. At the time-reversal invariant points, thanks to Eq.~\eqref{trasimm:eqn}, the Majorana correlators
\begin{eqnarray} \label{corrpm:eqn}
  &&\mean{(\opcdag{l}\pm\opc{l})(\opcdag{m}\pm\opc{m})}  \\
  &&= \pm \delta_{l\,m} - \frac{2i}{\pi}\int_0^\pi\Real(u_{k\,P}^-(t_b)(v_{k\,P}^-)^*(t_b))\sin(k(l-m))\ud k\nonumber
\end{eqnarray}
vanish for any $l\neq m$ (we have performed the thermodynamic limit $\frac{1}{L}\sum_{k>0}\to \frac{1}{2\pi}\int_0^\pi$, which is a very good approximation
in the case of $L\gg 1$). In this case the spin correlators
can be written as Toeplitz determinants of fermionic correlation matrices, which can be easily implemented numerically\cite{Barouch_PRA71}. Exploiting the translation invariance we can write
\begin{eqnarray} \label{correlators:eqn}
  \mean{\hat{\sigma}_j^z\hat{\sigma}_{j+l}^z}&=&\left|\begin{array}{cccc}G_{-1}&G_{-2}&\cdots&G_{-l}\\
                                                                       G_{0}&G_{-1}&\cdots&G_{-l+1}\\
                                                                       \vdots&\vdots&\vdots&\vdots\\
                                                                       G_{l-2}&G_{l-3}&\cdots&G_{-1}\end{array}\right|\nonumber\\
\quad\nonumber\\
  \mean{\hat{\sigma}_j^y\hat{\sigma}_{j+l}^y}&=&\left|\begin{array}{cccc}G_{1}&G_{2}&\cdots&G_{l}\\
                                                                       G_{0}&G_{1}&\cdots&G_{l-1}\\
                                                                       \vdots&\vdots&\vdots&\vdots\\
                                                                       G_{-l+2}&G_{-l+3}&\cdots&G_{1}\end{array}\right|
\end{eqnarray}
where we have defined the Majorana correlator
\begin{eqnarray} \label{corrb:eqn}
  G_{m-l}&=&\mean{(\opcdag{l}-\opc{l})(\opcdag{m}+\opc{m})}\nonumber\\
  &=&
%\frac{2}{L}\sum_{k>0}^{\rm ABC}
\frac{1}{\pi}\int_0^\pi\big[-2\Aimag(u_{k\,P}^-(t_b)(v_{k\,P}^-)^*(t_b))\sin(k(l-m))\nonumber\\
 &+&(|u_{k\,P}^-(t_b)|^2-|v_{k\,P}^-(t_b)|^2)\cos(k(l-m))\big]\ud k\,.
\end{eqnarray}
Similar expressions can be obtained for $\mean{\hat{\mu}_j^y\hat{\mu}_{j+l-1}^y}$, $\mean{\hat{\mu}_j^z\hat{\mu}_{j+l-1}^z}$ and $\mean{\hat{\beta}_j^y\hat{\beta}_{j+l-1}^y}$
applying the appropriate Jordan-Wigner
transformations. In the numerics, we construct the matrices of Eqs.~\eqref{correlators:eqn} using the mesh in $k$ given by the antiperiodic boundary conditions;
to evaluate the integrals~\eqref{corrb:eqn}, we use a cubic spline (see for instance~\ocite{NumericalRecipes}) and the integration routines of the {\em QUADPACK} package~\cite{quadpack:book}. %When we need to evaluate the
%correlators 
%far from the time-reversal invariant points
When Eq.~\eqref{trasimm:eqn} is not valid, we have to evaluate the Pfaffian of a matrix containing also the Majorana correlators Eq.~\eqref{corrpm:eqn} (all the formulae are given in detail in Ref.~\ocite{Barouch_PRA71}). To numerically evaluate these Majorana correlators we do in the same way as for those in Eq.~\eqref{corrb:eqn} and we compute the Pfaffians using the routines
of the {\em pfapack} library~\cite{pfapack}.

%In the next section we are going to show the correspondence between the order parameter in the spin representation at the time-reversal invariant point and the winding number in the fermionic representation. %{\bf Inserire discussione sulla simmetria $\mathbb{Z}_2$, l'invariante $\nu$ e il calcolo dei correlatori
%quando si \`e fuori dai punti ove c'\`e simmetria per inversione temporale.}
%

%
%We do not clearly understand these two-lobe structures.
%
%................................................................. Phase diagram .....................................................................%
%'''''''''''''''''''''''''''''''''''''''''''''''''''''''''''''''''''''''''''''''''''''''''''''''''''''''''''''''''''''''''''''''''''''''''''''''''''''%
\section{Phase diagram at a time-reversal invariant point} \label{orpam:sec}
%
%In this paper we take a 
We consider the system Eq.~\eqref{h11} undergoing a periodically driven field of the form
\begin{equation}
  h(t)=\left\{\begin{array}{ccc}h_0+A&\quad{\rm for}&\quad t\in[n \tau,(n+1/2)\tau]\\
                           h_0-A&\quad{\rm for}&\quad t\in[(n+1/2) \tau,(n+1)\tau]\end{array}\right.\,
\end{equation}
(see Fig.~\ref{fig:driving}). For concreteness we arbitrarily set $h_0=2.3$ and explore the phase diagram modifying $A$ and $\Omega$. %Thanks to the analytical insights provided in this paper, 
As we have remarked above, our analysis can be straightforwardly extended to any periodic drive.
To identify the phases of the FGS, we probe the value of the order parameters using Eqs.~\eqref{local_par:eqn},~\eqref{local_stag:eqn},~\eqref{Obeta:eqn},~\eqref{string_params:eqn},~\eqref{correlators:eqn} at the time-reversal invariant point $t_b=3\tau/4$.
%at times where there is no time-reversal invariance we use Eqs.~\eqref{local_par:eqn},~\eqref{local_stag:eqn},~\eqref{string_params:eqn} and the formulae
%discussed in Appendix~\ref{app_pfaff}. In all the cases, o
As we have discussed in the previous section, each order parameter is the limit of a specific spin correlator when the distance $l$ between the considered sites
diverges ($l\to\infty$ in Eqs.~\eqref{local_par:eqn},~\eqref{local_stag:eqn},~\eqref{Obeta:eqn},~\eqref{string_params:eqn}): 
we approximate these limit values 
%for $l\to\infty$ 
with the value of the correlators at finite $l\gg 1$. In all our calculations we take $L$ very large ($\sim 5000$) and we always consider $l<L/2$: in this range of $l$ the approximating correlators scale to their $l\to\infty$ limit as they would do in the thermodynamic limit.

Computing the spin order parameters at different points in the $(\Omega,A/\Omega)$ plane%, both in presence or absence of time-reversal symmetry
, we find many
different phases: in each phase only one order parameter is present while the others asymptotically tend to 0 for $l\to\infty$. The different phases correspond to
different colours in Fig.~\ref{diag_fase:fig}: we find that each phase transition corresponds to a degeneracy in the Floquet spectrum, in strict analogy to the case of standard equilibrium quantum phase transitions, which occur when the gap in the Hamiltonian closes~\cite{Sachdev:book}. Our model can be mapped to an integrable fermionic model, as we have discussed in Sec.~\ref{model:sec}. Evaluating the fermionic topological order parameter $w$, we find a precise correspondence between the nonvanishing string/local spin order parameter of each phase and the winding number $w$. This correspondence is synthetically shown in the table
on the right of Fig.~\ref{diag_fase:fig} and, for the values $w=0,\pm1,2$ coincides with the already-known results for a static system \cite{Manisha_PRB13}. We emphasize that the correspondence
among $O^\beta$ and $w=-2$ is a new finding of this work. We have additionally found phases with larger winding numbers, whose corresponding spin order is still unknown. (See in particular the phase diagram of Fig.~\ref{diag_fase:fig} which include phases with winding numbers $w=3$ and 4.)

Let us now comment on the structure of the phase diagram. The vertical transition lines occur at 
frequencies $\Omega_q = 2|h_0-1|/q$ (the ``$q$-series'') and $\Omega_p = 2|h_0+1|/p$ (the ``$p$-series) with $p,q$ integer numbers. 
These frequencies correspond to many-photon resonances of the periodically driven system, and are signaled by degeneracies of the Floquet spectrum respectively at the center of the Brillouin zone ($\mu=0$), or at its edge ($\mu=\Omega/2$). (See also Refs.~\ocite{Emanuele_arXiv15,russomanno_JSTAT15,angelo_arXiv16} for more details.) 

Let us consider a fixed $\Omega$ %between two consecutive resonances of this type and we 
and increase the amplitude $A$ of the driving. %as the amplitude $A$ of the driving is increased. As we can see, 
We notice, first of all, that the limit $A\to 0$ is in general singular. It is regular only if there are no resonances in the unperturbed spectrum in the limit $A\to 0$ (in our case, for $\Omega>\Omega_{p=1}=6.6$). In the presence of resonances, even a very small $A$ opens gaps in the spectrum and the topology
of the bands completely changes~\cite{Russomanno_JSTAT13}. Taking finite values of $A$, we can
see a whole series of quantum phase transitions (see Fig.~\ref{diag_fase:fig})
Interestingly, %we choose $\Omega$ between the two consecutive resonances, %in each of the strips separated by two
%consecutive vertical lines, 
the value of $A/\Omega$ of these transitions is almost independent of $\Omega$. % if we take as our variable . We notice that %-- for large
%$A$ -- these phases
%alternate periodically when $A/\Omega$ increases. For instance, for $\Omega_{p=2}<\Omega<\Omega_{p=1}$, $\Omega_{p=3}<\Omega<\Omega_{q=1}$, 
%$\Omega_{p=4}<\Omega<\Omega_{p=3}$, the different phases repeat themselves with a periodicity 2; on the opposite for 
%$\Omega_{q=1}<\Omega<\Omega_{p=2}$ the periodicity is $3.5$. In particular, we can see that 
The phase transitions approximately occur at $\frac{A}{\Omega}=\frac{j}{2}$ 
(for some integer $j$) and this approximation becomes better as $A/\Omega$ gets larger. As explained in appendix~\ref{resonances:sec}, this observation can be analytically justified by studying the properties of the Floquet Hamiltonian in an extended Hilbert space. This argument is analogous to the rotating wave approximation (RWA): we move to a time-dependent reference frame, where we apply perturbation theory and find that, when
$\frac{A}{\Omega}=\frac{j}{2}$, the Floquet Hamiltonian shows a degeneracy up to terms of order $(\Omega/A)^2$. See also Ref.~\ocite{Batisdas_PRA12} for a similar analysis of a sinusoidal time dependence of the driving, where the resonances are found at the zeros of Bessel functions.

\begin{figure}
  \begin{center} 
    \begin{tabular}{c}
%
%      \hspace{0.5cm}\\
      \resizebox{80mm}{!}{\includegraphics{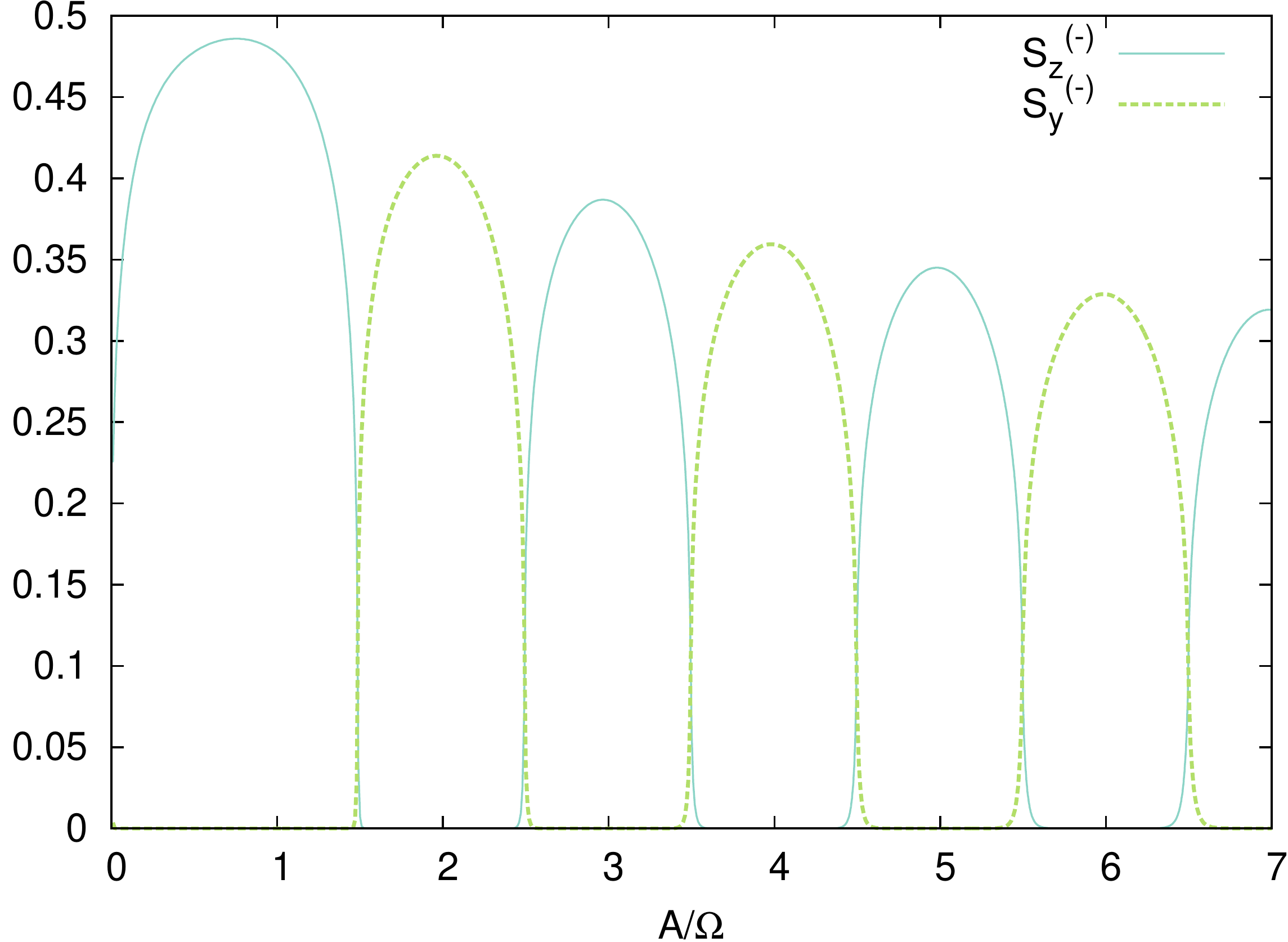}}\\
%      \hspace{0.5cm}\\
%      \resizebox{80mm}{!}{\includegraphics{carrellong_q2-p4-crop.pdf}}\\
%      \resizebox{80mm}{!}{\includegraphics{carrellong_p4-p5_VERO-crop.pdf}}
%
    \end{tabular}
  \end{center}
\caption{Local order parameters along the line $\Omega=5$ (with $h_0=2.3$). We can see that there is alternatively local order along $y$ (order parameter $S_y^{\,(-)}$)
and the $z$ (order parameter $S_z^{\,(-)}$). We show results for $L=4800$ and we approximate the order parameters (see Eq.~\eqref{local_par:eqn} and \eqref{local_stag:eqn}) with the value of the corresponding correlator for $l=200$. 
%Doing a finite size scaling we see that, in the thermodynamic limit, the derivative in $A/\Omega$ of the order parameters is discontinuous: the transition is second order. 
}
\label{local_fase:fig}
\end{figure}

\begin{figure}
  \begin{center} 
    \begin{tabular}{c}
%
%      \hspace{0.5cm}\\
      \resizebox{80mm}{!}{\includegraphics{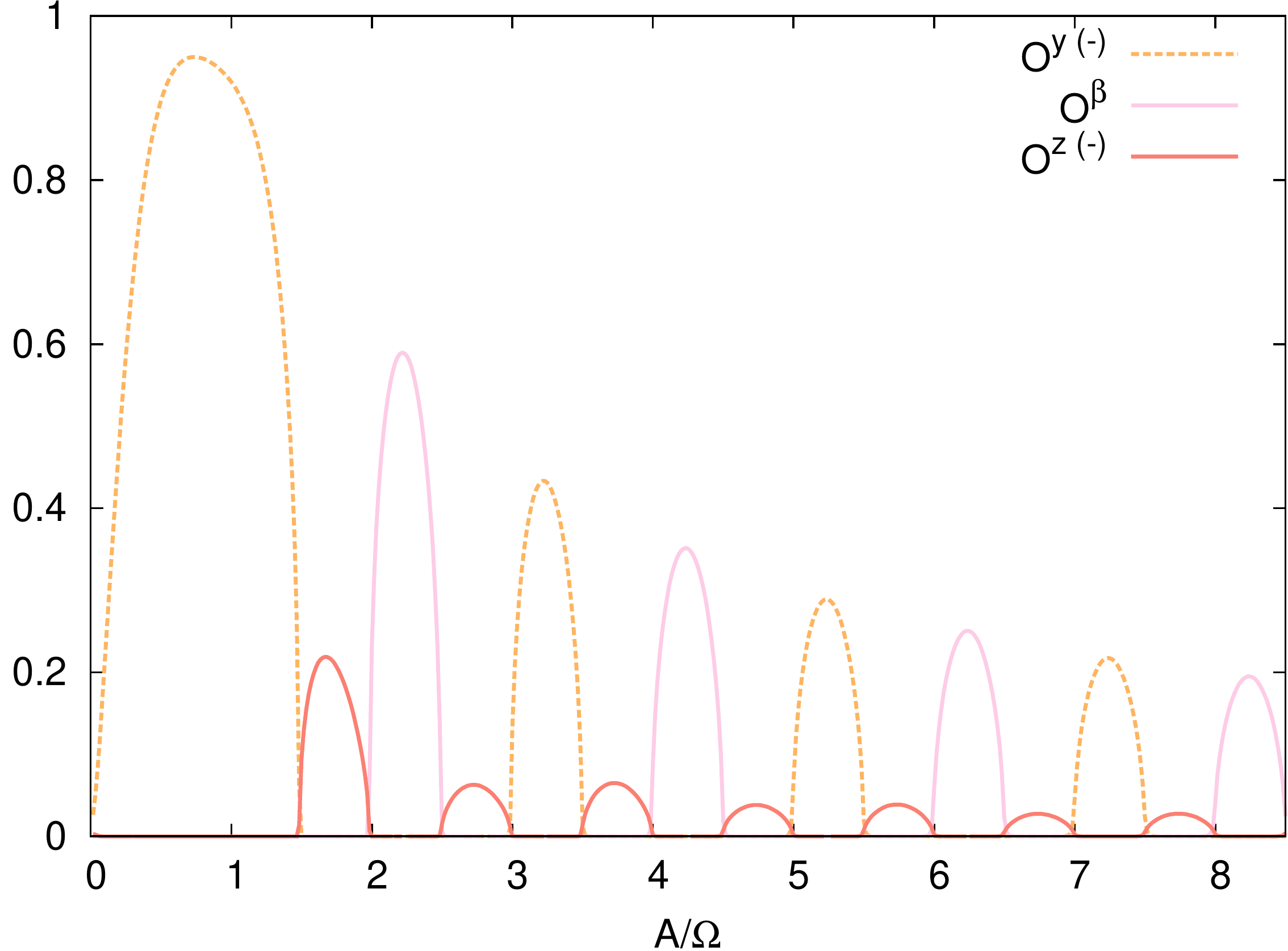}}\\
%      \hspace{0.5cm}\\
%      \resizebox{80mm}{!}{\includegraphics{carrellong_p4-p5-crop.pdf}}\\
%
    \end{tabular}
  \end{center}
\caption{String order parameters along the line $\Omega=3$ (with $h_0=2.3$). We can see that there is always string order: order parameters $O^y$,
$O^z$ and $O^\beta$ alternate with each other. We take $L=2400$ and we approximate the order parameters with the value of the corresponding correlator 
for finite $l=200$ 
(see Eqs.~\eqref{string_params:eqn},~\eqref{Obeta:eqn}). }%Similarly to Fig.~\ref{local_fase:fig}, we see that the transitions are second order in the thermodynamic limit.}
\label{string_fase:fig}
\end{figure}

\begin{figure}
  \begin{center} 
    \begin{tabular}{c}
%
%      \hspace{0.5cm}\\
      \resizebox{80mm}{!}{\includegraphics{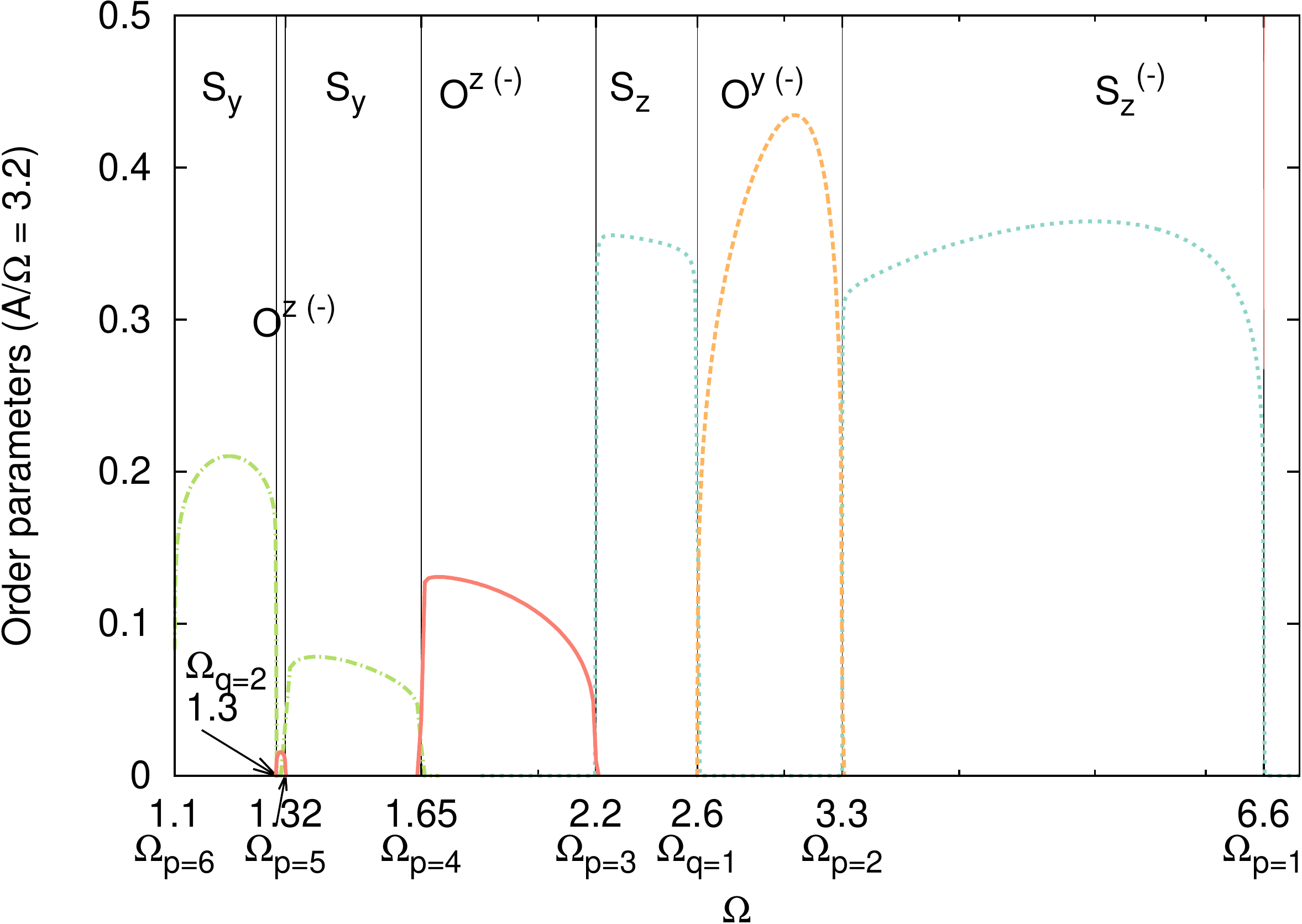}}\\
%      \resizebox{80mm}{!}{\includegraphics{carrellong_q2-p6-crop.pdf}}\\
%
    \end{tabular}
  \end{center}
\caption{Local and nonlocal order parameters along the line $A/\Omega=3.2$ (with $h_0=2.3$). The transitions are second order in the thermodynamic limit; we used $L=4800$ and we approximated the order parameters with the value of the corresponding correlator for $l=400$.}
\label{molte_fasi:fig}
\end{figure}

We now turn to the behavior of the order parameters at the transitions among the different phases %giving examples for some value of $\Omega$ in each strip. 
We specifically consider three ``cuts'' of the phase diagram shown in Fig.~\ref{diag_fase:fig}, referring to $t_b=3\tau/4$: (i) the vertical line at fixed $\Omega = 5.0$ (Fig.~\ref{local_fase:fig}), (ii) the vertical 
line at fixed $\Omega = 3.0$ (Fig.~\ref{string_fase:fig}),  (iii) the horizontal line at fixed $A/\Omega=3.2$ (Fig.~\ref{molte_fasi:fig}). In Fig.~\ref{local_fase:fig}, all the phases that we intersect possess local order parameters, while in Fig.~\ref{string_fase:fig} they possess string order parameters. In both cases we observe a series of lobes as a function of $A/\Omega$. The maximal height of each lobe decreases as $A$ is increased, suggesting that for $A\to\infty$ the system undergoes a complete mixing and all order parameters vanish. We have verified that this decrease occurs as a power law in $A$. In Figs.~\ref{local_fase:fig},~\ref{string_fase:fig} and~\ref{molte_fasi:fig} we show the finite-$l$ approximants of the order parameters: we see that they undergo some crossovers at the resonance points. These crossovers fully develop in phase transitions only in
the thermodynamic limit: for $l\to\infty$ the approximants tend to a finite value only in the phases where the corresponding
order parameters are nonvanishing, otherwise they scale to 0 (see Fig.~\ref{fs_scaling:fig} for an example).

For $l$ finite we see in the plots a sort of crossovers between the order parameters at the resonance points. The crossovers fully develop in phase transitions in the thermodynamic limit: for $l\to\infty$ the order parameter correlators tend to a finite value only in the phases where they are nonvanishing, otherwise they scale to 0 (see Fig.~\ref{fs_scaling:fig}(a) for an example).

%..........................%
%...........................%
%...........................%

%Emanuele: Ho cancellato questa frase perche' non presentiamo i calcoli numerici che supportano questo risultato 
%We have checked that  the decrease of the height is consistent with a power-law decay in $A$ ($h_{\rm max}\sim A^{-\gamma}$ with $\gamma>0$). %; the values of $\gamma$ we obtain

%CRITICAL EXPONENTS
From the dependence of the order parameters on the driving amplitude and frequency we can estimate the critical exponents of the transitions. In the present case, these calculations are simplified by our exact knowledge of the positions of the transitions (which correspond to the resonance points $\Omega_p$, $\Omega_q$ -- see above). In Fig.~\ref{fs_scaling:fig}(b) we consider the behavior of the
derivative $\ud S_z^{\,(-)}/\ud \Omega$ around the transition at $\Omega_{p=2}=3.3$. We see how the approximation for finite $l$ ($\ud S_z^{\,(-)}(l) / \ud \Omega$) shows a cusp that becomes higher with increasing $l$. In the limit of $l\to\infty$ it will give rise to a divergence, showing that the transition is of second order. %By performing a finite-size scaling (inset) w
We find that the local order parameter $S_z^{\,(-)}$ has a critical exponent $1/8$ at the transition $\Omega_{p=1}$ (see the inset in Fig.~\ref{fs_scaling:fig}): this is the same
critical exponent of $S_z$ in the static Ising transition. In contrast~\cite{Nota_scal}, the string-order parameter $O^y$ shows a critical exponent $0.25$ at the transitions $\Omega_{q=1}$ and $\Omega_{p=2}$. To understand this discrepancy we note that the string-order parameter corresponds to the limit of a correlator (of the dual model), while the local order parameters were defined as the square-root of the correlator. Furthermore we find that at the transitions, the quasi-energy of the FGS ($\overline{\mu}_{\rm FGS}=-\sum_{k>0}\mu_k$) shows a logarithmic divergences in its second derivative $\ud^2\mu/\ud\Omega^2$. This behavior is again analogous to the Ising transition of the static case, where the second derivative (with respect to the field) of the ground-state energy diverges logarithmically.
%In these plots we have considered the time-reversal invariant point $t_a=\tau/4$.
%

\begin{figure}
  \begin{center} 
    \begin{tabular}{c}
      \resizebox{80mm}{!}{\includegraphics{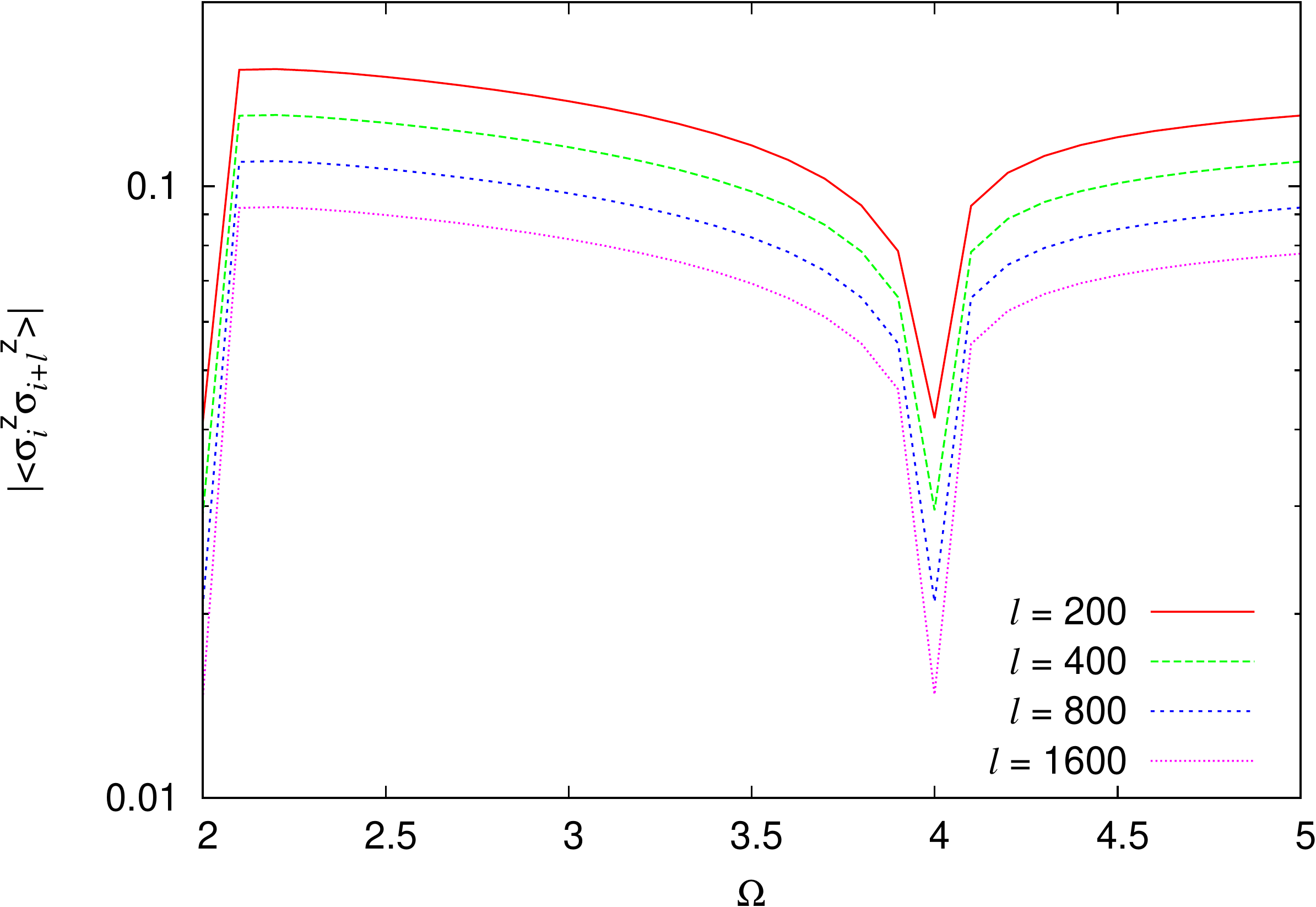}}
    \end{tabular}
  \end{center}
\caption{$|\mean{\hat{\sigma}_j^z\hat{\sigma}_{j+l}^z}|$ in the FGS versus $\Omega$ on the critical surface $h_0=1$, for different values of $l$. The power-law decay in $l$ reflects in the logarithm of $|\mean{\hat{\sigma}_j^z\hat{\sigma}_{j+l}^z}|$ decreasing of a constant amount when $l$ is doubled. Numerical parameters: $h_0=1,~A=0.3\,\Omega,~L=4800$.}
\label{scaling_critico:fig}
\end{figure}

We now briefly comment on the behavior of the Floquet ground mode at $\delta t$s that are not time-reversal invariant. In Fig.~\ref{local_fase_ntr:fig} we show the values of the order parameters $S^{(-)}_z$ and $S^{(-)}_y$ as a function of $\delta t$, for the same numerical parameters as in Fig.~\ref{local_fase:fig}. We find that for all $\delta t$, both order parameters vanish at the transition points. This finding is in agreement with the observation that the position of the phase transition does not depend on $\delta t$. However, the two above-mentioned order parameters can be used to characterize the different phases only at the time-reversal invariant points $\delta t=\tau/4,~3\tau/4$. For all other measurement times, both order parameters acquire a finite value in all phases. 
We can interpret this apparently anomalous situation in this way: the system shows long range order and the
order parameter is the staggered magnetization along a direction forming an angle $\alpha=\atan\left[S^{(-)}_z/S^{(-)}_y\right]$ with the $y$ axis; the modulus
of the order parameter is $\sqrt{{S^{(-)}_z}^2+{S^{(-)}_y}^2}$. (Notice that the symmetry of the Hamiltonian prevents the build up of 
mixed correlations of the form $\mean{ \hat{\sigma}_j^z \hat{\sigma}_j^y}$ in the eigenstates of the dynamics.)
Therefore, we can see that the order parameter rotates in time around the $x$ axis.
%This observation highlights the importance of the time-reversal invariant points in the characterization of the phases. In fact, as shown in Ref.~\ocite{Manisha_PRB13}, the existence of time-reversal invariant points is necessary for the distinction between the topological phases. 

\begin{figure}
  \begin{center} 
    \begin{tabular}{c}
      (a)\\
      \resizebox{80mm}{!}{\includegraphics{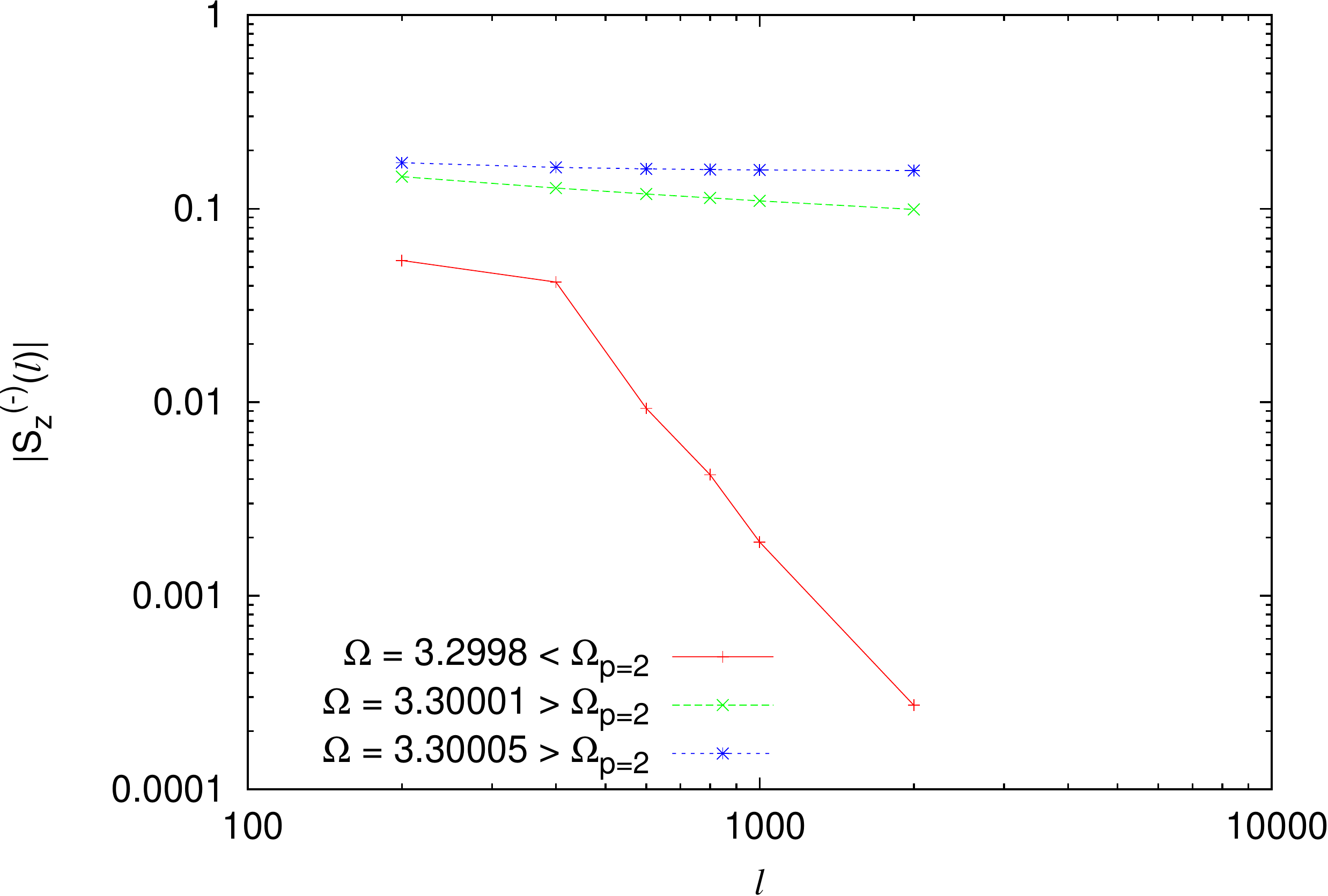}}\\
      (b)\\
      \resizebox{80mm}{!}{\includegraphics{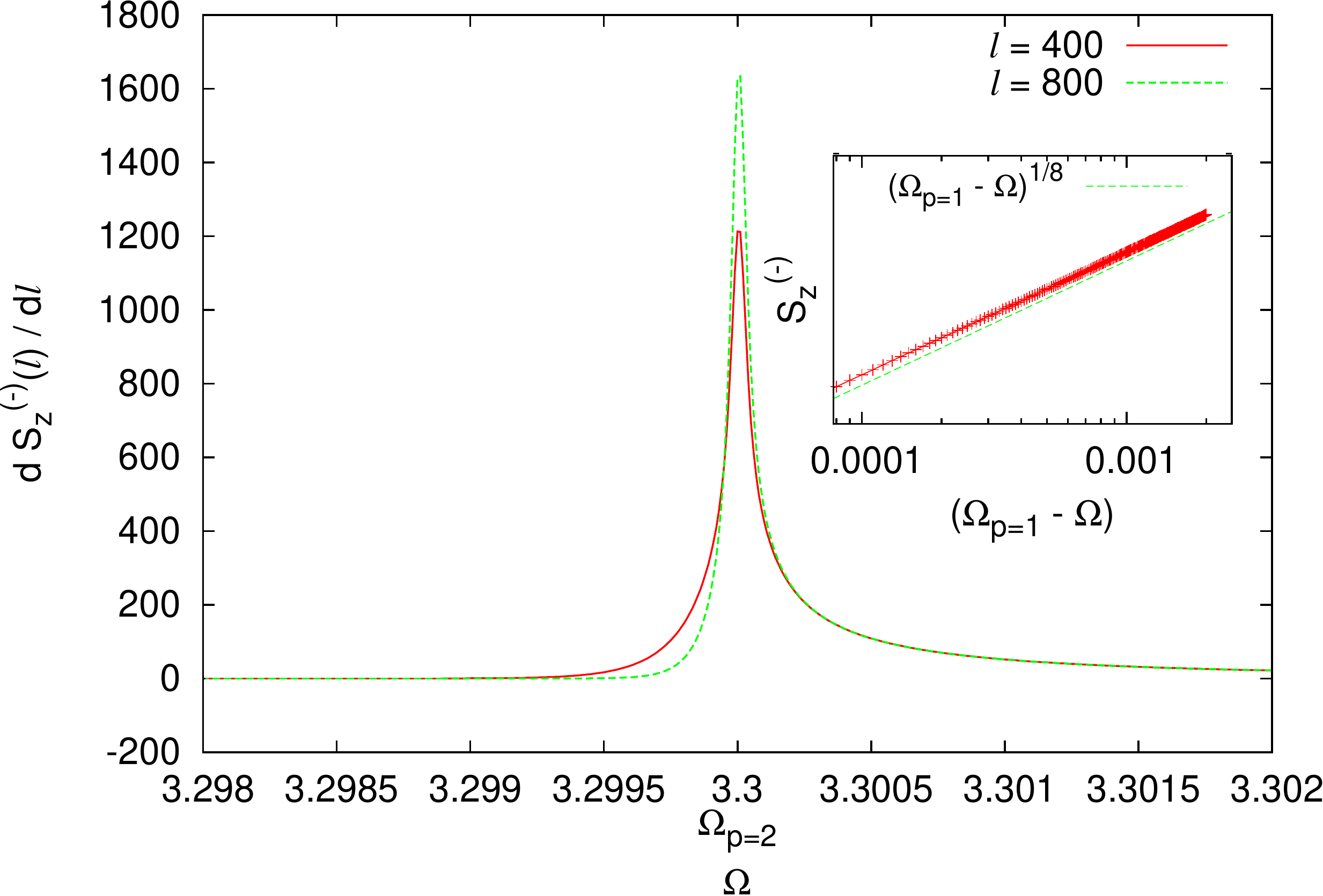}}
    \end{tabular}
  \end{center}
\caption{ (a) Scaling of $S_z^{\,(-)}(l)$ with $l$ close to the transition point. This object scales to 0 in a phase where the order parameter vanishes ($\Omega<\Omega_{p=2}=3.3$) and, on the opposite,
tends to a nonvanishing value in a phase with $S_z^{\,(-)}\neq 0$ ($\Omega>\Omega_{p=2}=3.3$). %: this limit is smaller when the transition is approached.
(b) Development of a divergence in $\ud S_z^{\,(-)}(l) / \ud \Omega$ in the limit $l\to\infty$. Numerical parameters: $h_0=2.3$, $A/\Omega=3.2$, and $L=4800$. (Inset) Scaling of the order parameter $S_z^{\,(-)}$
around the transition $\Omega_{p=1}$ in Fig.~\ref{molte_fasi:fig} (Numerical parameters: $L=4800,~l=800$).}
\label{fs_scaling:fig}
\end{figure}

We conclude this Section by considering the generalization of the previous findings for $h_0\neq2.3$. In the limit of large $\Omega$ the Floquet Hamiltonian is simply given by the time-averaged Hamiltonian, and corresponds to the static Ising model with $h\equiv h_0$. Thus, for $h_0>1$ the FGS is in the paramagnetic phase and shows $O^z$ order. In contrast, for $h_0<1$ the system is ferromagnetic and $S^z$ acquires a finite expectation value. This properties would hold for all frequencies $\Omega>2|1+h_0|$, where the first quantum phase transition occurs. The case of $h_0=1$ is special. For this value of the magnetic field the correlators decay to zero and there is no long range order. For each value of $A$ and $\Omega$, there is one of the order parameters of the phases with $h_0<1$ and $h_0>1$ such that the corresponding correlator decays algebraically for $h_0=1$, highlighting the quantum critical nature of this point. This behavior is exemplified in Fig.~\ref{scaling_critico:fig} where we show the correlator $|\mean{\hat{\sigma}^z_j\hat{\sigma}^z_{j+l}}|$ versus $\Omega$ for fixed $A$ and different values of $l$. Thanks to the logarithmic scale, we can see that the logarithm of the correlator
decreases of a constant quantity when $l$ is doubled: this reflects the power-law decay in $l$. 

\begin{figure}
\centering
\includegraphics[scale=0.6]{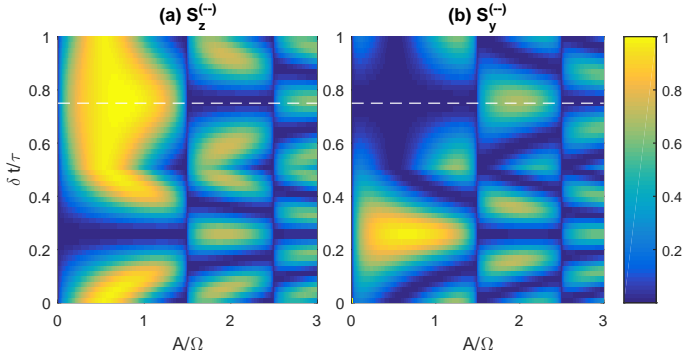} 
\caption{Order parameters $S_z^{(-)}$ and $S_y^{(-)}$ as a function of the driving amplitude $A$ and the measurement time $\delta t$. The dashed line highlights the time-reversal invariant point $\delta t=3\tau/4$, whose order parameters are shown in Fig.~\ref{local_fase:fig} and fully characterize the different phases. Note that at $\delta t=\tau/4$ the roles of the order parameters is interchanged. Numerical parameters: $h_0=2.3,~L=4800,~\Omega=5.0,~l=400$.}
\label{local_fase_ntr:fig}
\end{figure}

%...............................................................................................................................................................%
\section{Protected edge states} \label{OBC:sec}
\begin{figure}
  \begin{center} 
    \begin{tabular}{c}
%
%      \hspace{0.5cm}\\
      \resizebox{80mm}{!}{\includegraphics{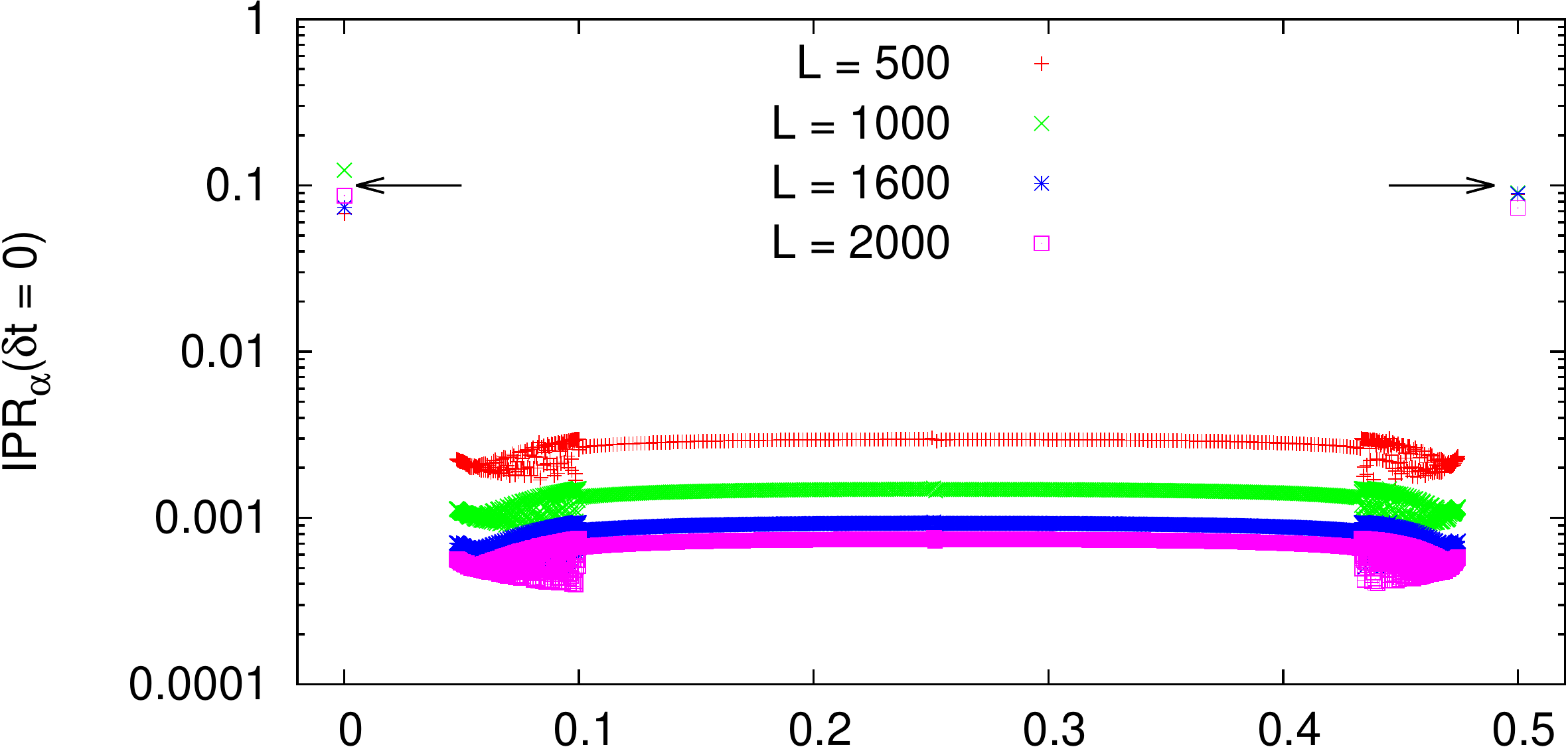}}\\
      \resizebox{80mm}{!}{\includegraphics{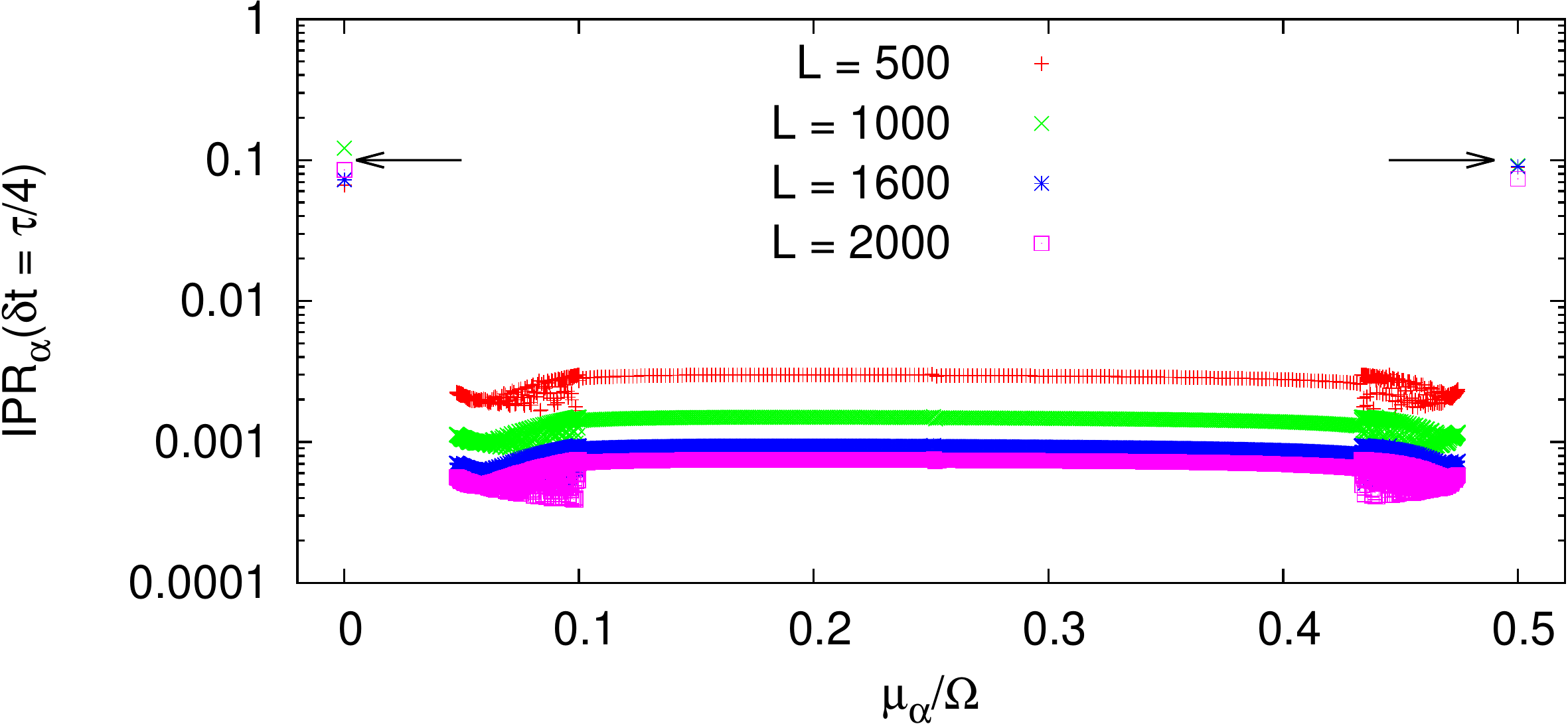}}\\
    \end{tabular}
  \end{center}
\caption{${\rm IPR}_\alpha(\delta t)$ vs. $\mu_\alpha$ for $\delta t=0$ in the top panel and $\delta t=\tau/4$ in the bottom panel (the IPR is in log scale). The arrows mark the two edge states at the edges of the quasi-energy spectrum: their IPR does not scale with $L$. The IPR of the bulk states in the bulk of the quasi-energy spectrum scales with $1/L$. This fact
is independent of $\delta t$. Numerical parameters: $h_0=2.3$, $\Omega=3.0$, $A=5.1$ ($\Omega$ falls in the second line of table~\ref{tab:edgemodes}, consistently we find an edge state at $\mu_\alpha=0$ and another one at $\mu_\alpha=\Omega/2$.)}
\label{ipr1:fig}
\end{figure}

When a system with nontrivial topology is put in contact with the vacuum,  zero-energy modes generically appear at the edge of the system~\cite{li2008entanglement,pollman2010entanglement}~\cite{Note3}. %\footnote{With the exception of topological phases that are protected by inversion symmetry, because this symmetry is broken  at the edges.} 
In one-dimensional topological superconductors the edge modes are zero-energy, topologically protected Majorana 
fermions~\cite{Alicea_RPP12}. Quite recently, also the case of periodically driven one-dimensional systems with $\mathbb{Z}_2$ symmetry has been considered (topological superconductors is a special case). Remarkably, single particle
Floquet edge modes have been discovered which can appear at quasi-energy 0 or $\Omega/2$~\cite{Liang_PRL06,Manisha_PRB13,Vedika_PRL16}. To detect the existence of edge modes in the Floquet Hamiltonian, we put the system in contact with the vacuum: from a technical point of view, this is equivalent to a chain with open boundary conditions. In this situation we cannot anymore apply the Fourier transform which led us to the simple 
expression for the Floquet Hamiltonian shown in Eq.~\eqref{HF:eqn}. Nevertheless, as we explain in appendix~\ref{Bogoliubov:sec}, we can define $L$ independent ``Floquet'' fermionic operators 
\begin{equation} \label{opazzi:eqn}
\opgamma{F,\alpha}(t)=\sum_{j=1}^L\left[{U}_{P\,j\alpha}^*(t)\opc{j}+{V}_{P\,j\alpha}^*(t)\opcdag{j}\right]\nep^{i\mu_\alpha t}
\end{equation}
(${U}_{P\,j\alpha}(t)$ and ${V}_{P\,j\alpha}(t)$ are $\tau$-periodic amplitudes), 
such that the Floquet Hamiltonian at time $\delta t\in[0,\tau]$ can be written as
\begin{equation} \label{generic_HF:eqn}
 \hat{H}_F(\delta t) = \sum_{\alpha = 1}^L\frac{\mu_\alpha}{2}
    \left(\opgammadag{F,\alpha}(\delta t)\opgamma{F,\alpha}(\delta t)-\opgamma{F,\alpha}(\delta t)\opgammadag{F,\alpha}(\delta t)\right)\,.
\end{equation}
The resulting Hamiltonian has a quadratic form with single-particle quasi-energies $\mu_\alpha$: each Floquet operator $\opgamma{F,\alpha}(\delta t)$
corresponds to a one-particle Floquet state of the system $\ket{\psi_\alpha(\delta t)}=\opgammadag{F,\alpha}(\delta t)\ket{0}$ with quasi-energy $\mu_\alpha$.
In appendix~\ref{Bogoliubov:sec} we elucidate how to numerically compute the single-particle quasi-energies and the amplitudes $\{{U}_{P\,j\alpha}(\delta t),\,{V}_{P\,j\alpha}(\delta t),\,j=1,\ldots,L\}$. To detect if among the single quasi-particle Floquet states there are some edge states, 
we exploit the fact that edge states are localized at the edge: their inverse participation ratio (IPR)~\cite{Edwards_JPC72} in physical space, defined as
\begin{equation}
  {\rm IPR}_\alpha(\delta t) = \sum_{j=1}^L\left(\left|{U}_{P\,j\alpha}(\delta t)\right|^4+\left|{V}_{P\,j\alpha}(\delta t)\right|^4\right)
\end{equation}
does not scale with $L$ in the limit of $L\gg 1$ (see also Ref.~\ocite{Manisha_PRB13}). On the opposite, being the system we are considering clean, the other single-particle Floquet states are extended, and their IPR scales as $\sim 1/L$. Using this method the fermionic edge states are not difficult to recognize. %{\bf \`E possibile far vedere una figura dell'IRP per stati localizzati e non localizzati?}
We can see an instance of this fact in Fig.~\ref{ipr1:fig}, where we plot ${\rm IPR}_\alpha(\delta t)$ versus $\mu_\alpha$ for different values of $\delta t$ and of $L$. In this figure we consider a case in which there are two single-particle Floquet states localized at the boundaries: they appear at the edges of the single-particle quasi-energy spectrum, one at $\mu_\alpha=0$ and the other at $\mu_\alpha=\Omega/2$. We can see that the IPR of the edge modes does not scale with $L$ and is much larger than the IPR of the ones in the bulk of the spectrum. The IPR of the other (bulk) states, on the contrary, scales as $1/L$. This can be clearly seen in the semi-logarithmic plot: doubling $L$, the logarithm of the IPR of these states decreases of a constant value. Moreover, we can see that this behaviour is independent of the considered time $\delta t$: the dynamics conserves the number of edge states in the FGS and their position in the spectrum.

\begin{table} %\label{w_e_Nd:tab}
 \begin{tabular}{|c|c|c|c|c|c|c|}
  \hline
  $\Omega$ & $w$ & $N_b$ & $N_b(0)$ & $N_b(\Omega/2)$ \\
  \hline
  %\hline  & $w=\pm$ 1 ($\nu=-1$) & $w=2$ ($\nu = 1$)\\
  \hline $\Omega>\Omega_{p=1}$               & 0 		 & 0 & 0 & 0 \\
  \hline $\Omega_{p=2}<\Omega<\Omega_{p=1}$  & $\pm 1$	 & 1 & 0 & 1 \\ 
  \hline $\Omega_{q=1}<\Omega<\Omega_{p=2}$  & 0,$\pm 2$ & 2 & 1 & 1 \\
  \hline $\Omega_{p=3}<\Omega<\Omega_{q=1}$  & $\pm 1$	 & 1 & 1 & 0  \\ 
  \hline $\Omega_{p=4}<\Omega<\Omega_{p=3}$  & 0,$\pm 2$ & 2 & 1 & 1 \\ 
  \hline $\Omega_{p=5}<\Omega<\Omega_{p=4}$  & $\pm 1$,3 & 3 & 2 & 1  \\ \hline
%  \hline$\Omega_{q=2}<\Omega<\Omega_{p=5}$  & +1 & 4 & 2 & 2 \\ \hline
%         Number of edge states &
 \end{tabular}
 \caption{Topological numbers characterizing the first six vertical strips in the phase diagram (see Fig.~\ref{diag_fase:fig}: the winding number $w$; the total number of edge states $N_b$; the number of edge states at $\mu_\alpha=0$ ($\mu_\alpha=\Omega/2$), $N_b(0)$  ($N_b(\Omega/2)$). The associated Floquet topological numbers~\cite{Roy_arXiv16} are respectively $n_C=w$ and $n_L=\pm N(\Omega/2)$(see text for details).}
\label{tab:edgemodes}
\end{table}

We find that the number of edge states $N_b$ is time-independent and constant within each phase, confirming its topological nature. As mentioned above, in contrast to equilibrium topological insulators, in Floquet topological phases edge states can occur both in the center of the Floquet band, $\mu_\alpha=0$, and at its edge, $\mu_\alpha=\Omega/2$, (see also Refs.~\ocite{Liang_PRL06,Manisha_PRB13,Vedika_PRL16}).
We empirically find that both $N_b(0)$ and $N_b(\Omega/2)$ do not depend on the driving amplitude $A$, and are unchanged on each of the six vertical strips of the phase diagram of Fig.~\ref{diag_fase:fig}. This empirical rule is only broken in the strip $\Omega_{q=2}<\Omega<\Omega_{p=5}$: in this frequency domain, $N_b(0)=2$ for  $A/\Omega<2.0$ and $A/\Omega>2.3$, and $N_b(0)=0$ for $2.0\,\Omega\lesssim A\lesssim 2.3\,\Omega$. This point deserves further investigation. In particular we will need to clarify whether the maximal system-size considered in our calculations, $L=2000$, is sufficient to reach the thermodynamic regime of the small intermediate phase (whose gap is small and correlation length very long).

\begin{figure}
\centering
\includegraphics[scale=0.7]{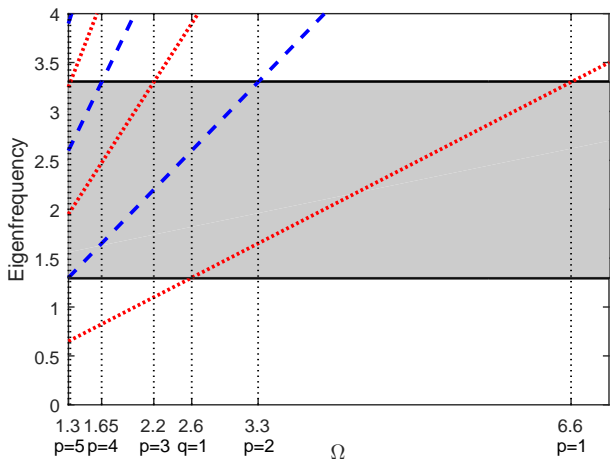}
\caption{Graphical method to determine the number of edge states as a function of the driving frequency\textcolor{black}{, adapted from Ref.~\ocite{Manisha_PRB13}} (see text for details). The gray area represents the positive band of the time-independent part of the Hamiltonian. The diagonal lines correspond to $\mu=n\Omega/2$ with $n\in{\mathcal N}$: red dotted lines represent odd $n$s, and blue dashed lines represent even $n$s. The intersections of the diagonal lines and the upper edge of the band correspond to the $p$-series resonances ($2|h_0+1|=p\Omega$), and the intersections with the lower edge of the band correspond to the $q$-series resonances ($2|h_0-1|=q\Omega$). For a given frequency $\Omega$, the number of red dotted (blue dashed) lines inside the gray area is equal to $N_b(0)$ ($N_b(\Omega/2)$). %The results obtained by this analysis are in agreement with the number of edge modes determined by the numerical solution of the model.
}
\label{fig:edgemodes}
\end{figure}

The number of edge states in each frequency domain is summarized in Table \ref{tab:edgemodes} and agrees with the arguments of Ref.~\ocite{Manisha_PRB13}. In particular, $N_b=0$ in the high-frequency phase, which is adiabatically connected to the static paramagnet. With decreasing frequency, $N_b$ increases by one at the resonances of the $p$ series ($\Omega_p=2|h_0+1|$) and decreases by one at the $q$ series ($\Omega_q=2|h_0-1|$)~\cite{Nota_closing}. The frequency of the added (subtracted) edge state depends on the parity of the $p$ ($q$) index: $N_b(0)$ increases (decreases) at even $p$s ($q$s), and $N_b(\Omega/2)$ increases (decreases) at odd $p$s ($q$s). A graphical representation of this result is shown in Fig.~\ref{fig:edgemodes}. 

In order to put our work in a broader perspective, we notice that our results are consistent with the classification of non-interacting topological Floquet systems presented in Ref.~\ocite{Roy_arXiv16}. The authors are able to decompose the evolution operator over one period of a driven system
in two parts: a unitary loop component and a constant evolution component. The topological numbers associated with these two components respectively characterize the topology of the evolution over one period (loop component, $n_L$) and of the Floquet Hamiltonian (constant component, $n_C$). Thanks to a homotopy equivalence they show that there exists only a finite number of symmetry classes for periodically driven integrable systems. We can see that our case, because of time inversion and particle hole symmetries, falls in the symmetry class \textcolor{black}{BDI} with $d=1$: this gives rise to a \textcolor{black}{classifying} group $\mathbb{Z}\times\mathbb{Z}$, where $n_C$ and $n_L$ are arbitrary integer numbers.

The quantum number $n_C$ classifies the topology of Floquet Hamiltonian and therefore equals to the winding number $w$ reported in Fig.~\ref{diag_fase:fig}, while $n_L$ is a distinct topological index. These two quantum numbers uniquely determine the number of edge modes in the system. According to the bulk-edge correspondence principle, the number of Majorana edge states equals to the absolute value of an associated bulk topological winding number: in our case $N_b(0) = |n_0|$ and $N_b(\Omega/2)=|n_\pi|$, where~\cite{Roy_arXiv16} $n_0=n_C+n_L=w+n_L$ and $n_\pi=n_L$. These relation imply for instance that $|w|$ is always smaller or equal than $N_b(0)+N_b(\Omega/2)$, in agreement with the findings of Table \ref{tab:edgemodes}. For concreteness, let us consider the three phases mentioned in its last row. These phases share the same $N_b(0)=2$ and $N_b(\Omega/2)=1$, but differ by their winding numbers, which equal to $w=+1$, $-1$, and $3$. The associated constant and loop topological numbers are respectively $(n_C,n_L) = (1,1),~(-1,-1),$ and $(3,-1)$.

\subsection{Time-translation symmetry breaking of the Floquet ground state}

\textcolor{black}{The presence of Majorana fermions with quasienergy $\Omega/2$ (i.e. $N(\Omega/2)\neq 0$) is associated with the FGS spontaneously breaking the discrete time-translation symmetry~\cite{Nayak_PRL16,zhang_16:preprint,choi_16:preprint,Vedika_PRL16,Vedika_PRB16,time_crystals_exp:preprint,Vedika_la_santa,moessner2017equilibration}. To understand this point, let us focus on the ferromagnetic phases obtained for $\Omega_{p=2}<\Omega<\Omega_{p=1}$ where $N(\Omega/2)=1$ (see Table \ref{tab:edgemodes}). Combining the two Majorana modes at the edges of the system, we can define a Fermionic operator at quasienergy $\Omega/2$: switching its occupation is equivalent to the addition of $\Omega/2$ to the many-body quasienergy. As a consequence, in this phase, each Floquet eigenstate has a partner at quasienergy difference $\Omega/2$.}

%{\bf Qua ci andrei cauto. Nel modello di Ising ci sono i Majorana modes a energia 0 ma solo il GS rompe la $\mathbb{Z}_2$ symmetry. Questo perch\'e gli stati eccitati sono massivamente degeneri. Il disordine rompe questa degenerazione e porta tutto lo spettro a rompere la $\mathbb{Z}_2$ symmetry nel caso statico e la TTS in quello dinamico. In questo caso si ha il time-crystal nel senso di Nayak. Restringerei perci\`o il discorso al solo FGS.}}  they $\mathbb{Z}_2$ 
%
{ In particular, by considering the even and odd superpositions of the FGS and its partner, we can construct a state which is periodic under $2\tau$~\cite{Nayak_PRL16}: this is possible thanks to the existence of global symmetry breaking and long range order along $y$ or $z$. The FGS and its partner are long range correlated~\cite{Nayak_PRL16}: in the spin basis they respectively correspond  to the symmetric and the anti-symmetric superposition of a symmetry breaking state with spin up and another with spin down. If we prepare the system in one of the two symmetry breaking states, we see that it is flipped to the symmetry-breaking state with opposite spin after each driving. This commutation occurs thanks to the quasi-energy difference of $\Omega/2$ of the two Floquet states whose superposition gives the state: this phenomenon is formally analogous to the Rabi oscillations. Looking at the order-parameter magnetization (along $z$ or $y$, according to the specific phase), we would see oscillations of period $2\tau$: spontaneously breaking the global spin symmetry is therefore essential to break the time-translation symmetry~\cite{moessner2017equilibration,Nayak_PRL16,Vedika_la_santa,Vedika_PRL16,Sondino_PRB16}.}

{ Like the excited states do in the static quantum Ising chain, the Floquet states different from the FGS do not break the spin symmetry because of the very non-local nature of the Floquet excitations; hence they cannot break even the time-translation symmetry. That's why our system is not a time-crystal: in order to see time-translation symmetry breaking for a very wide class of initial states, all the Floquet spectrum (or at least an extensive fraction of it) must break the time-translation symmetry~\cite{Nayak_PRL16,arXiv:io}. {On the opposite, in our case only preparing the
system in one of the symmetry breaking Floquet Ground states would allow the observation of the period doubling oscillations}. This result is in agreement with the findings of Ref.~\ocite{Vedika_la_santa} which rule out the existence of time-crystals for clean systems with short range interactions.} %{\bf Questo legame \`e alla base del fatto che gli stati eccitati non rompono la TTSB}.}

{In order to give a further connection with known systems displaying the time-translation symmetry breaking, we see that the phases in $\Omega_{p=2}<\Omega<\Omega_{p=1}$ are equivalent to the $\pi$ ferromagnet described in Ref.~\ocite{Vedika_PRL16}, respectively in the $z$ and the $y$ directions. Similarly, the phases obtained for $\Omega_{p=2}<\Omega<\Omega_{q=1}$ with $w=0$ correspond to the $0\pi$ phase of Ref.~\ocite{Vedika_PRL16}.
%~\footnote{In this phase $N(0)=N(\Omega/2)=1$ and one can construct a fermion from the superposition of two Majorana modes at the same edge of the sample, with quasienergy difference $\Omega/2$. This state has periodicity $2\tau$: upon Jordan-Wigner transformation it is mapped to a rotating spin edge state.}.
} 
%{\bf There being {\em one single} pair of states state showing time-translational symmetry breaking, this system is not a Floquet time-crystal~\cite{Nayak_PRL16}: only a very specific preparation of the initial state would allow the observation of the period doubling oscillations. This result is in agreement with the findings of Ref.~\ocite{Vedika_la_santa} which rule out the existence of time-crystals for clean systems with short range interactions.}}

%
%.............................................................................................................................................%
\section{Scaling of the correlation length in a ramping} \label{ramping_correlation:sec}
%
%It would be nice to experimentally prepare the FGS and explore this rich phase diagaram, but -- as we have alredy pointed out -- there are many difficulties. 
From an experimental perspective, it is necessary to specify a protocol to prepare the Floquet ground state. One possibility would be to prepare the ground state of the time-averaged Hamiltonian at high frequency and then reach the FGS by adiabatically lowering the frequency~\cite{Emanuele_arXiv15}. This is possible if no
resonance in the Floquet spectrum is met~\cite{Messiah:book,Breuer_ZPD89,Young_70,Polko_adiab}. But if we want to prepare, for instance, a FGS with $S_z^{(-)}\neq 0$ starting from infinite frequency, we have to cross
a resonance (the one at $\Omega=\Omega_{p=1}$). 
In crossing a resonance, the adiabatic theorem does not apply and it is not possible to exactly prepare the FGS. In fact, one can show that the order parameter
remains exactly zero across the transition. Nevertheless, 
if we change our Hamiltonian very slowly, we can obtain a state with an arbitrary long %$S_z^{(-)}\neq 0$ 
correlation length but never infinite. We numerically estimate the $S_z^{(-)}$ correlation length as
\begin{equation}
 l_z = \frac{\sum_{j=0}^{l_{\rm max}}j\left|\mean{\hat{\sigma}_1^z\hat{\sigma}_{1+j}^z}\right|}
           {\sum_{j=0}^{l_{\rm max}}\left|\mean{\hat{\sigma}_1^z\hat{\sigma}_{1+j}^z}\right|}
\end{equation}
where we choose $l_{\rm max}$ so that the sum has already reached convergence. We ramp the frequency in time as 
$$
  \Omega(t)=\Omega_i+\frac{1}{t_f}(\Omega_f-\Omega_i)\,;
$$
we take $h_0=2.3$, $\Omega_i=10.0$, $\Omega_f=4.0$, $A=1.0$; $t_f$ is the characteristic time scale of the ramping. 
We ramp the system from a phase with order $O^z$ to a phase with order $S_z^{(-)}$ at the time-reversal
invariant point $t_b=3\tau/4$ (see Fig.~\ref{diag_fase:fig}).
The driving with linearly changing frequency is a square wave of the form $h_0+A\sign(\sin(\phi(t)))$ with $\phi(t)=\Omega(t)t$: the crossing of the resonance occurs 
when $d\phi/d t = \Omega_{p=1}$. We probe the system at times $t_n$ such that $\phi(t_n)=2n\pi+\frac{3}{2}\pi$; in this way we can make comparison with the system at the time reversal invariant point $t_b=3\tau/4$~\cite{Nota_infty}. Because this protocol breaks the time reversal invariance, the Majorana correlator Eq.~\eqref{corrpm:eqn} is nonvanishing and we need to use the general formula involving the Pfaffian to evaluate the correlator (see Sec.~\ref{parametra:sec}).

\begin{figure}
  \begin{center} 
    \begin{tabular}{c}
%
%      \hspace{0.5cm}\\
%      \resizebox{80mm}{!}{\includegraphics{carrellong_p2-q2-crop.pdf}}\\
%      \hspace{0.5cm}\\
      \resizebox{80mm}{!}{\includegraphics{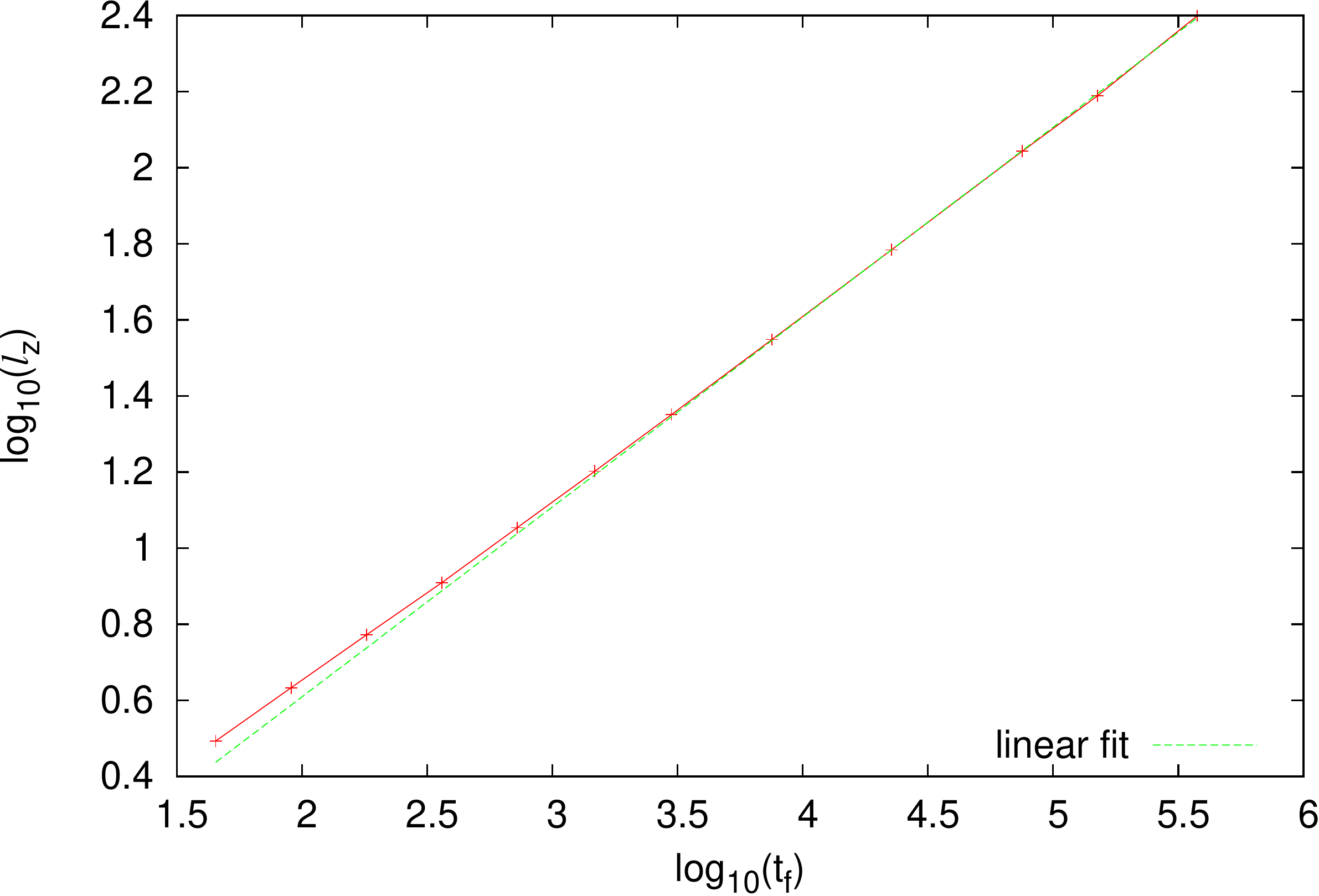}}\\
      \resizebox{80mm}{!}{\includegraphics{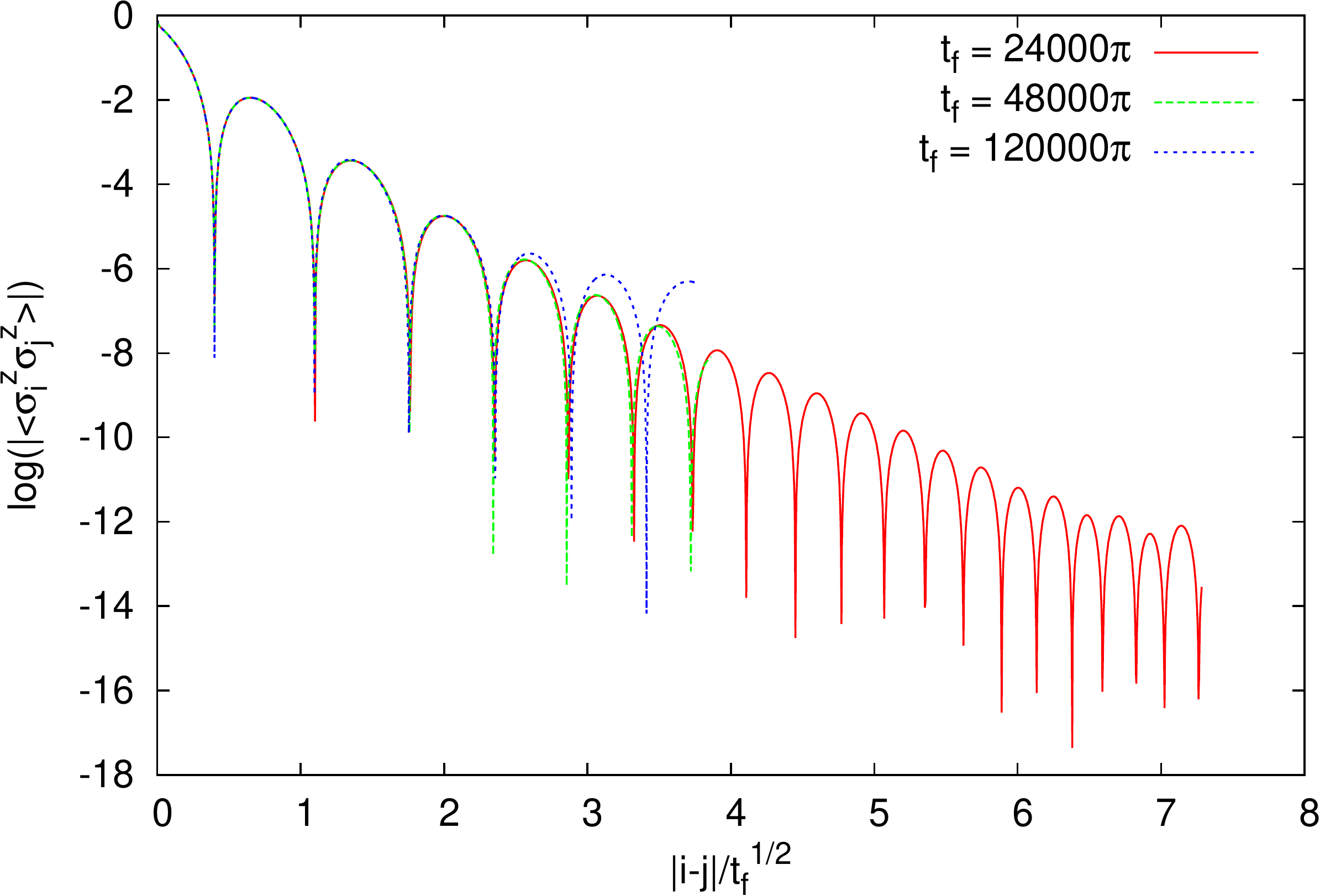}}
    \end{tabular}
  \end{center}
\caption{(Upper panel) Scaling of the correlation length $l_z$ at the end of the frequency ramping, with the ramping time $t_f$ in a bilogarithmic plot (red curve). The steepness of
the red curve in figure increases with $t_f$ converging asymptotically to a constant: for large $t_f$, $l_z$ scales polynomially with $t_f$: $l_z\propto t_f^\alpha$.
 The linear fit in the bilogarithmic plot, performed for $\log(t_f)>3.5$, provides $\alpha=0.499\pm 0.003$. (Lower panel) Logarithm of the correlator $|\mean{\hat{\sigma}_i^z\hat{\sigma}_j^z}|$ at the end of the ramping vs. $|i-j|$ rescaled by $t_f^{1/2}$ for different values of $t_f$. We see that all the curves have the same steepness: the correlation length rescaled by $t_f^{1/2}$ is independent of $t_f$. The calculations were performed on a chain of size $L=3800$.}
\label{inlen1:fig}
\end{figure}

In analogy to the standard Kibble-Zurek phenomenon~\cite{Zerek_PRA07}, we find that the correlation length at the end of the ramping, $l_z$, scales polynomially with  $t_f$: $l_z\sim t_f^{\alpha}$. We show this scaling in the upper panel of Fig.~\ref{inlen1:fig} by means of a bilogarithmic plot. We find that the steepness of the curve $\log_{10}(l_z)$ versus $\log_{10}(t_f)$ is an
increasing function which saturates to an asymptotic value when $t_f\to\infty$. %asymptotically we have $$. 
By means of a linear fit of this curve in the region of large $t_f$ we numerically find $\log_{10}(l_z)=\alpha\log_{10} t_f + {\rm const}$ with $\alpha= 0.499\pm 0.003$. In our numerical calculations $t_f\leq120000\pi$ and the steepness has reached a satisfying convergence: %We therefore conjecture that the analytically-exact value of the critical exponent is 
our result is consistent with $\alpha=0.5$, in agreement with the scaling of the defect density found in Ref.~\ocite{Emanuele_arXiv15}, $n_{\rm ex}\sim1/\l_z\sim t_f^{-0.5}$. In the lower panel of the figure we plot the correlator $\log|\mean{\hat{\sigma}_i^z\hat{\sigma}_j^z}|$ versus $|i-j|$ rescaled by $t_f^{1/2}$: we see
that all the rescaled curves have the same decay length and this confirms the scaling of $l_z$ with $t_f^{1/2}$.

%Emanuele: Ho cancellato queste due frasi perche' interrompono il filo del discorso e non mi sembra che aggiungano nuove importanti informazioni - mi sembra che le figure + caption siano "self-explanatory".

%To make this point even more explicit, in the lower panel of Fig.~\ref{inlen1:fig} we plot $\log|\mean{\hat{\sigma}_i^z\hat{\sigma}_j^z}|$ versus $|i-j|$ rescaled by $t_f^{0.492}$. First of all we see that$|\mean{\hat{\sigma}_i^z\hat{\sigma}_j^z}|$ always shows an exponential decay (which appears linear in the single logarithmic plot). Second, we see that with this rescaling all the lines have the same steepness: the correlation length rescaled by $t_f^{0.492}$ is independent of $t_f$. }
%for $A=1$: %and we ramp the frequency in time as $\Omega(t)=\Omega_i+(\Omega_f-\Omega_i)/t_f$, 
%The scaling exponent of the correlation length after the ramping is given by $z\nu/(1+z\nu)$, where $-z$ is the scaling exponent of the correlation length with
%$\Omega-\Omega_{p=1}$ at the transition and $\nu$ is the scaling exponent of the gap. For this transition we find numerically $z=1$ and $\nu=1$, so the scaling exponent we find is consistent with the static scaling exponents.
%*we find that after the ramping the correlation length of the correlator $(-1)^l\mean{\sigma_j^z\sigma_{j+l}^z}$ scales with $t_f^{1/2}$ (see Fig.~\ref{inlen1:fig}); this is consistent with the transition at $\Omega_{p=1}$ being of the second order. Moreover, the scaling exponent we find is consistent with the static scaling exponents we find at that transition.
%

\begin{figure}
  \begin{center} 
    \begin{tabular}{c}
%
%      \hspace{0.5cm}\\
%      \resizebox{80mm}{!}{\includegraphics{carrellong_p2-q2-crop.pdf}}\\
%      \hspace{0.5cm}\\
      \resizebox{80mm}{!}{\includegraphics{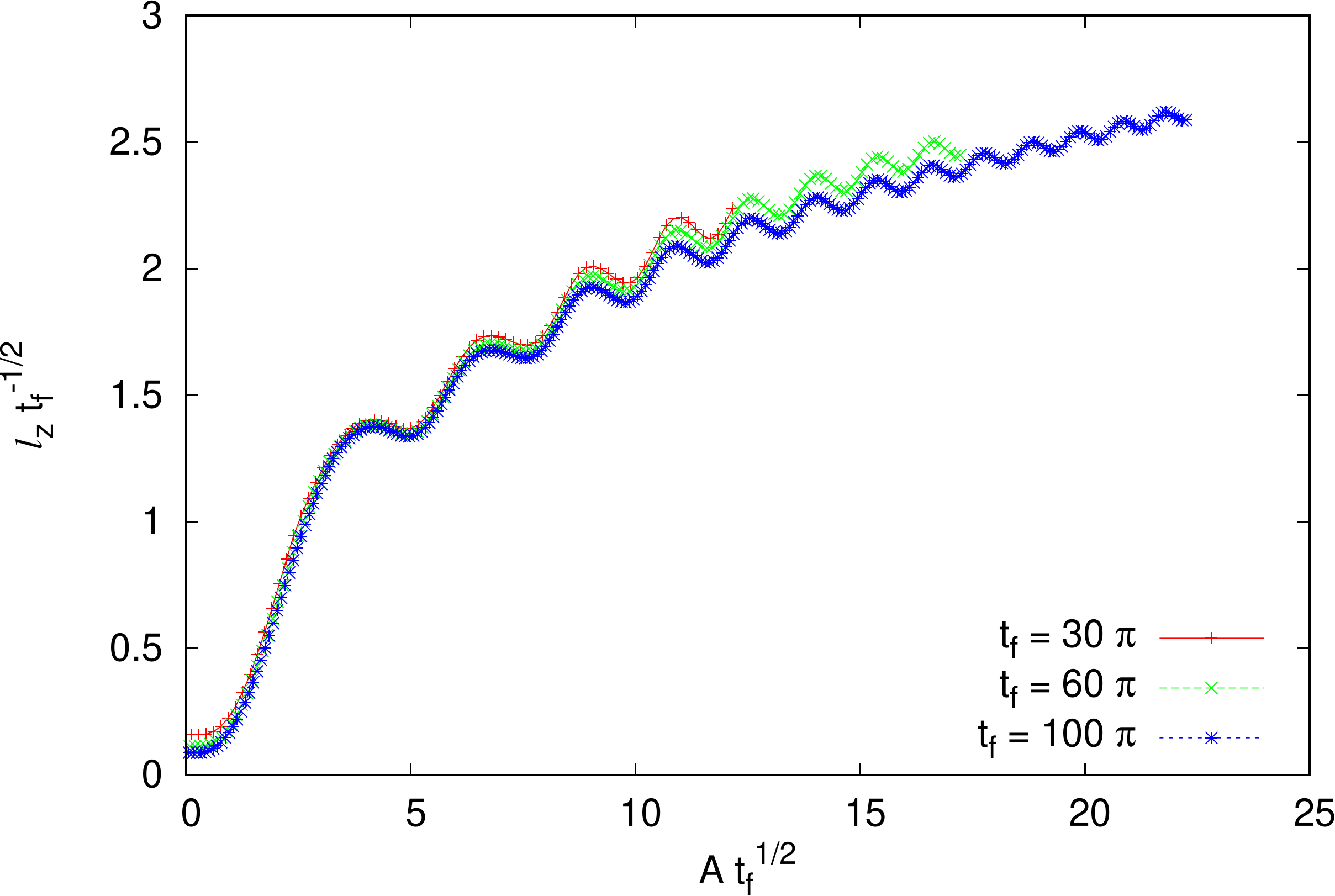}}\\
    \end{tabular}
  \end{center}
\caption{ Rescaled $l_z t_f^{-1/2}$ vs. $A t_f^{1/2}$: we see that the curves for different $t_f$ collapse. Numerical parameters: $L=5000,~h_0=2.3,~A_f=1.0,~\Omega=5.0$.}
\label{inlen2:fig}
\end{figure}

An alternative method to prepare the system in a nontrivial FGS is to fix the frequency and adiabatically increase the amplitude of the driving: $A(t) = A_f t / t_f$. In this case we probe the correlators at times $3n\tau/4$, with $n$ integer. We set $\Omega=5.0$
and $A=1.0$ and perform an adiabatic ramp inside a phase with order parameter $S_z^{\,(-)}$ (see Fig.~\ref{diag_fase:fig}).  Note that in this case our starting point ($A=0$) is a critical point of the Floquet spectrum where the Floquet gap is closed. As a consequence, the validity of the adiabatic theorem is not guaranteed \cite{polkovnikov2008breakdown}. We nevertheless find a very clear scaling of the correlation length, as in the standard crossing of the quantum critical point: actually the correlation length behaves as a scaling function: $l_{z}\sim t_f^{1/2}\mathcal{L}(At_f^{1/2})$ (see Fig.~\ref{inlen2:fig}).

We conclude discussing the effective possibility to prepare the Floquet ground state. The protocol we describe is analogous to the
one used in the standard Kibble-Zurek case. If the system is in the thermodynamic limit, it always arrives to a state which is not the final Floquet ground
state, but an approximation to it, as better as the ramping is slower (the Floquet gap closes at the critical point and
we are never exactly adiabatic). In particular,
the range of the correlations will be longer when the ramping is slower. If, after the
ramping, the stroboscopic Floquet Hamiltonian is kept static (the frequency is unchanged), the system will slowly evolve towards an
asymptotic condition; because the system is integrable, the asymptotic condition is given by a Floquet generalized Gibbs ensemble
~\cite{Russomanno_PRL12,Lazarides_PRL14,angelo_arXiv16} (ergodic systems evolve instead towards a $T=\infty$ thermal ensemble~\cite{Rigol_PRX14,Lazarides_PRE14,Ponte_AP15,Rosch_PRA15,Russomanno_EPL15}). %Generalized Gibbs ensemble if the system is integrable 
%or a thermal ensemble if it is ergodic. 
The transient is as longer as the final state is
nearer to the Floquet ground state. %Here occurs exactly the same, being the driven Ising model integrable,
%it will evolve towards a non-thermal Floquet generalized Gibbs ensemble~\cite{Russomanno_PRL12,Lazarides_PRL14,angelo_arXiv16}.
So, the slower we perform the ramping, the nearer the state is to the Floquet
ground state, the longer it survives (and can be observed) before there is relaxation towards the GGE.

Moreover, if we consider a finite system, the gap at the quantum critical point is nonvanishing. If we perform the
ramping slowly enough, we can be exactly adiabatic and prepare the exact Floquet ground state. 
Another possibility, which will be the object of future work, is to couple the ramped driven system to a
noise able to make it to relax to the Floquet ground state (this has been already done
for systems undergoing a ramping of the static Hamiltonian~\cite{Che_ciolla_PRL17}).
%................................................. CONCLUSION .................................................................%
\section{Discussion}\label{concludo:sec}
In this work we have studied the phase diagram of the Floquet ground state of a periodically driven integrable quantum spin chain. Its quantum
phase transitions occur at well-defined resonances in the Floquet spectrum, whose position is here approximated by analytical expressions and confirmed by numerically-exact calculations. Thanks to the Jordan-Wigner transformation, the problem can be mapped to a chain of non-interacting fermions. In the case of time-reversal invariant protocols, this fermionic model includes an infinite number of distinct topological phases, characterized by a topological winding number~\cite{Manisha_PRB13,Emanuele_arXiv15}. Here we show that in the original spin representation the phases can be characterized by a set of local and nonlocal (string) order parameters, such that a single nonvanishing long-ranged order corresponds to each phase. At present, we are able to determine the order parameters for all the phases with winding numbers $w=0~,\pm1,~\pm2$. In particular, the phase with $w=-2$ is characterized by a previously unnoticed string order parameter, $O^\beta$, introduced in Eq.~(\ref{Obeta:eqn}). Based on our findings, we conjecture that phases with odd (even) winding numbers are characterized by local (nonlocal) order parameters of increasing complexity.  We have furthermore showed some cases where the correlation length associated to the order parameter shows a Kibble-Zurek scaling when slowly crossing one of the non-equilibrium quantum phase transitions. 

As predicted by Ref.~\ocite{Roy_arXiv16}, we find that the Floquet quantum phases can be characterized by two topological numbers $n_0$ and $n_\pi$, whose absolute value respectively gives the number of edge states at quasi-energies $0$ and $\Omega/2$. The sum of these two numbers equals the winding number of the Floquet Hamiltonian.  %Interestingly, the number of protected edge states is invariant in time, while %both the fermionic and the bosonic 
%the order parameters are uniquely defined at time-reversal-invariant points only. 
{{Four of the phases that we describe in our clean non-interacting model are adiabatically connected to the four phases of the clean non-interacting version of the kicked Ising model considered in Fig.2(a) of Ref.~\ocite{Vedika_PRL16}. The comparison is meaningful, showing both models time-reversal invariant points and being both models clean and integrable through the Jordan-Wigner transformation (we do not consider here the more complicated disordered and interacting version of the model in Ref.~\ocite{Vedika_PRL16}). {The correspondence of the phases is as follows: }the PM phase found by those authors corresponds to our paramagnetic phases with $w=0$ and no edge mode; their ferromagnet to our ferromagnetic phases with $w=1$ and the edge state at $\mu=0$; their $\pi$-ferromagnet to our ferromagnetic phases with $w=1$ and the edge state at $\mu=\Omega/2$; and their $0\pi$ phase to our phases with $w=0$ and edge states at $0$ and $\Omega/2$}. The model considered by these authors has additional {symmetries (duality and symmetries under discrete translation/inversion of some driving parameters)} that prevent it from entering the entire range of phases described in this paper. The situation is analogous to the static Ising model in transverse field (Eq.~\eqref{h11} with $h(t)=h_0$): although the model belongs to the BDI class with $\mathbb{Z}$ classification group, this model actually shows two phases (paramagnet and ferromagnet) only.}

%
%Our work \textcolor{black}{addresses} fundamental questions about the classification of topological phases in periodically driven one-dimensional fermionic systems, and in particular about the relation between winding numbers and edge states. It additionally This result highlights the nontrivial relation between these topological quantities and the local and nonlocal orders of the  unitarily equivalent spin model. 
%Perspectives of future work concern the study of the phase diagram in the disordered driven system, especially in connection with topological properties (quantum phase transitions in the static disordered system are already well known~\cite{Fisher_PRB95}). 
%Considering also terms that break integrability, it would be interesting to study the existence of topological order in all the eigenstates of the dynamics in case of many body localization~\cite{Chittarospo_JSTAT13,disko_stu_prb13,paraj2016anomalous,Vedika_PRL16,norman_16:preprint}. 

\textcolor{black}{Perspectives for future work concern the effects of integrability breaking terms, that are mapped under Jordan-Wigner to interactions among the fermions. These terms have two important effects: first, in analogy to the equilibrium case, interactions are known to restrict the size of the classification group. In the case of periodically driven interacting fermions with parity and time-reversal symmetry, the relevant classification group is $\mathbb{Z}_8\times \mathbb{Z}_2\times \mathbb{Z}_2$~\cite{Sondina_PRB16}. A second important effect deals with the nature of the Floquet ground state: for any finite-size system, this state can always be defined by %considering a finite-size system, 
starting from a macroscopically large driving frequency, and adiabatically reducing it. Will the state obtained by this procedure hold the same properties as true ground states (of time independent Hamiltonians)?}

\textcolor{black}{The common wisdom is that this is not the case: generic non-integrable systems are expected to thermalize and have Floquet eigenstates with volume law entanglement. Under this assumption, the Floquet ground state would not show any phase transition and would be always topologically trivial. Two possible workarounds that have been discussed in the literature are {\em (i)} using disorder and many-body localization to prevent heating and thus keep a low entanglement entropy (see for example Refs.~\ocite{Ponte_PRL15,moessner2017equilibration} for an introduction), {\em (ii)} { perform the adiabatic following at a small but finite rate in order to achieve a transient low-entanglement state before the system reaches the thermal state}~\cite{Buco_arXiv15}.
%to skip over ``mini-gaps'' and prevent thermalization to higher Floquet bands 
 Another option that one may wish to consider is the possibility that Floquet ground states do not thermalize and can show low entanglement. This question is correlated to the problem of stability of classical many-body systems (see for example Ref.~\ocite{Emanuele_2014:preprint}) and deserves further investigation.}

\begin{acknowledgements}
We acknowledge useful discussions with E. Berg, E. Demler, J. Goold, R. Fazio, F.~P. Landes, N. Lindner, A. Polkovnikov, L. Privitera, D. Rossini, G.~E. Santoro and F.~M. Surace. We are indebted with
G.~E.~Santoro for the subroutine performing the simultaneous diagonalization discussed in Appendix~\ref{Bogoliubov:sec}. This work was supported in part by the Israeli Science Foundation Grant No. 1542/14. A.~R. acknowledges financial support from the EU integrated project QUIC, from his parents %, from UNICREDIT  
and from "Progetti interni  -  SNS". Part of the numerical calculations were performed on the {\em Goldrake} cluster in Scuola Normale Superiore.
This work was not supported by any military agency. 
\end{acknowledgements}
\appendix
%.......................................................................................................................................................%
%
\section{Symmetries of the Floquet Hamiltonian} \label{time-rev:app}
To obtain the Floquet Hamiltonian of a mode $k$ ($\mathbb{H}_{k\,F}(\delta t)$ -- Eq.~\eqref{HF:eqn}) we need to solve the Bogoliubov-de Gennes equations 
(Eq.~\eqref{deGennes:eqn}) over one period of the driving. In this way we can construct the time-evolution operator over one period
\begin{equation} \label{Ut:eqn}
  \mathbb{U}_k(\tau+\delta t,\delta t) = \overleftarrow{\mathcal{T}}\nep^{-i\int_{\delta t}^{\tau+\delta t}\mathbb{H}_k(t')\ud t'}\,,
\end{equation}
where $\overleftarrow{\mathcal{T}}$ is the time-ordering operator; the Floquet Hamiltonian is obtained as
\begin{equation} \label{UktF:eqn}
  \mathbb{U}_k(\tau+\delta t,\delta t) = \nep^{-i\tau\mathbb{H}_{k\,F}(\delta t)}\,.
\end{equation}
Using these relations, we can obtain the symmetries of $\mathbb{H}_{k\,F}(\delta t)$ starting from those of $\mathbb{H}_k(t)$.

First of all, at all times the system is invariant under particle-hole symmetry: $\tau_x\mathbb{H}_k(t)\tau_x = - \mathbb{H}^*_{-k}(t)$ -- see Eq.~\eqref{tras1:eqn};
we can find indeed~\cite{note_roy}
\begin{equation}
  \tau_x\mathbb{U}_k(\tau+\delta t,\delta t)\tau_x = \overleftarrow{\mathcal{T}}\nep^{-i\int_{\delta t}^{\tau+\delta t}\tau_x\mathbb{H}_k(t')\tau_x\ud t'} = \mathbb{U}_{-k}^*(\tau+\delta t,\delta t)\,.
\end{equation}
Using Eq.~\eqref{UktF:eqn} we can see that $\tau_x\mathbb{H}^*_{k\,F}(\delta t)\tau_x = - \mathbb{H}_{-k\,F}(\delta t)$: the Floquet Hamiltonian is particle-hole symmetric whatever the value of $\delta t$.

Let's consider a time $t_r$ where the system is time-reversal invariant; thanks to the periodicity of the driving, 
time reversal invariance will repeat with period $\tau$. The time-dependent Hamiltonian indeed enjoys the symmetry relation
\begin{equation}
  \mathbb{H}_k(t_r+n\tau+t) = \mathbb{H}^*_{-k}(t_r+n\tau-t)\,.
\end{equation}
Using this relation in Eq.~\eqref{Ut:eqn} we can find
\begin{equation} \label{UktF1:eqn}
  \mathbb{U}_k(\tau+t_r,t_r) = \overleftarrow{\mathcal{T}}\nep^{-i\int_{0}^{\tau}\mathbb{H}^*_{-k}(t_r-t')\ud t'}\,.
\end{equation}
With an appropriate change of variables and exploiting the $\tau$-periodicity of $\mathbb{H}_k(t)$ we can write
\begin{equation} \label{UktF2:eqn}
  \mathbb{U}_k(\tau+t_r,t_r) = \overleftarrow{\mathcal{T}}\nep^{-i\int_{t_r}^{t_r+\tau}\mathbb{H}^*_{-k}(t'')\ud t''} = \left(\mathbb{U}_{-k}^\dagger(\tau+t_r,t_r)\right)^*\,.
\end{equation}
Using Eq.~\eqref{UktF2:eqn}, we find $\mathbb{H}_{k\,F}(t_r)=\mathbb{H}^*_{-k\,F}(t_r)$: the Floquet Hamiltonian enjoys time-reversal symmetry at the
time-reversal invariant points.

We have indeed shown that the symmetry relation~\eqref{tras2:eqn} are valid for the Floquet Hamiltonian at the time-reversal invariant
points. We conclude this Section showing how the relations Eqs.~\eqref{timer:eqn} and~\eqref{trasimm:eqn} can be derived. We focus on the case of the Floquet
Hamiltonian, the case of the static Hamiltonian is exactly the same. In general, we can write the Floquet Hamiltonian as
\begin{equation}
 \mathbb{H}_{k\,F}(t_r) = A_k\tau_x+B_k\tau_y+C_k\tau_z\,,
\end{equation}
where we have defined the real quantities
\begin{eqnarray}
  A_k&=&-2\mu_k\Real\left(u_{k\,P}^-\left(t_r\right)\left(v_{k\,P}^-\left(t_r\right)\right)^*\right)\,,\nonumber\\
  B_k&=&-2\mu_k\Aimag\left(u_{k\,P}^-\left(t_r\right)\left(v_{k\,P}^-\left(t_r\right)\right)^*\right)\,,\nonumber\\
  C_k&=&\mu_k\left(\left|v_{k\,P}^-\left(t_r\right)\right|^2-\left|u_{k\,P}^-\left(t_r\right)\right|^2\right)\,.
\end{eqnarray}
Applying the time-reversal symmetry (Eq.~\eqref{tras2:eqn}) we find
\begin{equation}
  A_k = A_{-k},\quad B_k= - B_{-k},\quad C_k = C_{-k}\,.
\end{equation}
Instead applying the particle-hole symmetry (Eq.~\eqref{tras1:eqn}) we obtain
\begin{equation}
  A_k = - A_{-k},\quad B_k= - B_{-k},\quad C_k = C_{-k}\,.
\end{equation}
The only way in which these two systems of equations can be both true is that $A_k=0$; in the case $\mu_k\neq 0$ this implies the conclusion Eq.~\eqref{trasimm:eqn}.
The quasi-energy $\mu_k$ can be vanishing only at isolated points, indeed Eq.~\eqref{trasimm:eqn} holds for almost every $k$ and this is enough to ensure the vanishing
of the integral in Eq.~\eqref{corrpm:eqn}.
%
%.......................................................................................................................................................%
\section{Resonances in the Floquet Hamiltonian} \label{resonances:sec}
As shown for instance in Refs.~\ocite{Shirley_PR65,Hausinger_PRA10}, finding the Floquet modes and quasi-energies amounts to diagonalize a static Hamiltonian in an infinite-dimensional Hilbert space. Expanding
the periodic Hamiltonian in Fourier series
\begin{equation}
  \hat{H}(t) = \sum_n \hat{H}^{(n)}\nep^{-in\Omega t}\,,
\end{equation}
we need to diagonalize the infinite matrix
\begin{equation}
  \left(\begin{array}{ccccccc}
                  \hat{H}^{(0)}+2\Omega       &  \vdots     &  \vdots     &  \vdots  &\vdots & \cdots   \\
           \cdots&\hat{H}^{(0)}+\Omega&\hat{H}^{(1)}&\hat{H}^{(2)}&\hat{H}^{(3)}&\cdots\\
           \cdots&\hat{H}^{(-1)}&\hat{H}^{(0)}&\hat{H}^{(1)}&\hat{H}^{(2)}&\cdots\\
           \cdots&\hat{H}^{(-2)}&\hat{H}^{(-1)}&\hat{H}^{(0)}-\Omega&\hat{H}^{(1)}&\cdots\\
           \cdots&\hat{H}^{(-3)}&\hat{H}^{(-2)}&\hat{H}^{(-1)}&\hat{H}^{(0)}-2\Omega&\cdots\\
           \end{array}\right)\,.
\end{equation}
We define this object as the Floquet extended Hamiltonian.
We see that the blocks on the diagonal are the $n=0$ Fourier components shifted by an integer number of $\Omega$. The $n$-th Fourier component is on the
$n$-th progressive diagonal of the matrix. We see that this Hamiltonian is invariant if we add $k\Omega\boldsymbol{1}$ (this is equivalent of applying to the
Floquet states a rotation of angle $\nep^{-ik\Omega t}$). Thanks to this symmetry, the Floquet spectrum results invariant under translations of an integer number
of $\Omega$, as it should be.

We apply this analysis to the $k$-th component of the single-particle Hamiltonian $\mathbb{H}_k(t)$ (see Eq.~\eqref{Ht:eqn}). Before doing that, we apply to this
matrix a unitary time-dependent transformation
\begin{equation}
  \mathbb{V}(t) = \nep^{i\phi(t)\sigma_z}
\end{equation}
where
\begin{eqnarray}
  \phi(t) &=& -\int_0^t (h(t')-h_0)\ud t' \\
          &=& -A\left\{\begin{array}{ccc}
                                                     t-k\tau&{\rm if}&k\tau<t<(k+1/2)\tau\\
                                                     t-(k+1/2)\tau&{\rm if}&(k+1/2)\tau<t<(k+1)\tau
                                                   \end{array}\right.\nonumber\,.
\end{eqnarray}
The Hamiltonian in the rotated frame is
\begin{eqnarray}
  \widetilde{\mathbb{H}}_k(t) &=& \mathbb{V}^\dagger(t)\mathbb{H}_k(t)\mathbb{V}(t)-i\mathbb{V}^\dagger(t)\dot{\mathbb{V}}(t) \nonumber\\
    &=& 
  \left(\begin{array}{cc}
     h_0-\cos k & -i\sin k \nep^{-2i\phi(t)}\\
     i\sin k \nep^{2i\phi(t)}&\cos k - h_0
  \end{array}\right)\,.
\end{eqnarray}
At this point we can evaluate the Fourier coefficients and construct the Floquet extended Hamiltonian. We find
{\small
\begin{eqnarray}
  \widetilde{\mathbb{H}}_k^{(0)} &=& \left(\begin{array}{cc}
     h_0-\cos k & -i\sin k \frac{1}{2\pi}\frac{\Omega}{A}\sin\left(2\pi\frac{A}{\Omega}\right)\\
     i\sin k \frac{1}{2\pi}\frac{\Omega}{A}\sin\left(2\pi\frac{A}{\Omega}\right)&\cos k - h_0
  \end{array}\right)\,,\nonumber\\
  \quad\nonumber\\
  \widetilde{\mathbb{H}}_k^{(n)} &=& \sigma_y\frac{\Omega}{\pi}\cos\left(\pi\frac{A}{\Omega}-n\frac{\pi}{2}\right)\nonumber\\
       &\times&\left[\frac{\sin\left(\pi\frac{A}{\Omega}-n\frac{\pi}{2}\right)}{A-n\frac{\Omega}{2}}+\frac{\sin\left(\pi\frac{A}{\Omega}+n\frac{\pi}{2}\right)}{A+n\frac{\Omega}{2}}\right]\,.\nonumber\\
\end{eqnarray}
}
First of all, we see that -- when $k=0$ or $k=\pi$ -- all the off-diagonal terms in the Floquet extended Hamiltonian vanish: our matrix is already diagonal and
we can easily see if there are resonances.
In the case $k=0$, $\widetilde{\mathbb{H}}_k^{(0)}=(h_0-1)\sigma_z$. Therefore -- if $h_0 -1 = -1 + h_0 + p\Omega$ -- one diagonal term of $\widetilde{\mathbb{H}}_k^{(0)}-n\Omega$
is equal to one diagonal term of $\widetilde{\mathbb{H}}_k^{(0)}-(n-p)\Omega$ and we have a Floquet resonance (we have called these resonances ``the $p$-series''). In the case
$k=\pi$, we find the resonance condition $2(h_0+1)=q\Omega$ and we have the resonances of the $q$-series. These resonances correspond to the vertical transition
lines of Fig.~\ref{diag_fase:fig}. 

Dealing with the horizontal transition lines, we have to discuss the off-diagonal term of $\widetilde{H}_k^{(n)}$. There are two cases. 
If $n=0$ or $n\geq 1$ is odd, the off-diagonal term vanishes when $A = \frac{j}{2}\Omega$, with $j$ even. If $n=0$ or $n\geq 1$ is even, the off-diagonal term vanishes when
$A=\frac{j}{2}\Omega$, with $j$ odd. When one of these two conditions is valid, then there are infinite pairs of diagonal terms of the Floquet extended Hamiltonian 
which are connected by a vanishing matrix element. For instance, if we have the condition for $j$ even, we have that $h_0-\cos k$ has vanishing matrix
element with $-h_0 +\cos k $, $-h_0 +\cos k + \Omega$, $-h_0 +\cos k + 3\Omega$ and so on. On the opposite, if we have the condition valid for $j$ odd, 
we have %the resonance for $k$ and $n$ satisfying $h_0=\cos k+n\Omega$ with $n$ even. 
that $h_0-\cos k$ has vanishing matrix
element with $-h_0 +\cos k $, $-h_0 +\cos k + 2\Omega$, $-h_0 +\cos k + 4\Omega$ and so on. If we apply perturbation theory (similarly to what done in Ref.~\ocite{Hausinger_PRA10}), we see that the corrections
to the two degenerate levels are second order in $(\Omega/A)$. So, up to second order in $\Omega/A$, we have degeneracies in the Floquet spectrum when $A=\frac{j}{2}\Omega$. Therefore, the approximation is better for large $A/\Omega$ and this is confirmed by the phase diagram in Fig.~\ref{diag_fase:fig}.

To conclude this Appendix, we say that our phase transitions in $A/\Omega$ closely remind those found in Ref.~\ocite{Batisdas_PRA12} for a quantum Ising chain with a sinusoidal
driving $h(t)=h_0+A\sin(\Omega t)$. Those transition points were obtained by means of a change to a rotating reference frame + the rotating wave approximation. This method is equivalent to apply the perturbative approximation we have discussed. If we apply the perturbation theory we have just discussed to the sinusoidally-driven model, we  obtain the same transition points of Ref.~\ocite{Batisdas_PRA12}, occurring at the zeros of the Bessel functions $J_l(2A/\Omega)$.
%
%..............................................................................................................................................................%
%\section{Correlators in absence of time-reversal symmetry}
%\label{app_pfaff}
%
%...............................................................................................................................................................%
\section{Floquet Hamiltonian for the non translationally invariant system} \label{Bogoliubov:sec}
%-----------------------------------------------------------------------------------------------------------%
%
In this Appendix we briefly describe the quantum dynamics of non translationally invariant Ising/XY 
chains~\cite{Caneva_PRB07,Russomanno_JSTAT13} undergoing a periodic driving of period $\tau$. We aim to obtain the general expression of the 
Floquet Hamiltonian Eq.~\eqref{generic_HF:eqn}.
%whenever there is no translational invariance and Fourier transform cannot be applied; 
We start introducing the Bogoliubov-de Gennes dynamics, closely following the discussion of Refs.~\ocite{Russomanno_JSTAT13,angelo_arXiv16}. 
Generically, if $\opc{j}$ denote the $L$ fermionic operators originating from the Jordan-Wigner transformation of spin operators~\cite{Lieb_AP61}
\begin{eqnarray}
  \label{wj}
  \hat{\sigma}_j^x&=&1-2\opcdag{j} \opc{j}\nonumber\\
  \hat{\sigma}_j^z&=&\hat{\tau}_j\left(\opcdag{j}+\opc{j}\right)\nonumber\\
  \hat{\sigma}_j^y&=&i\hat{\tau}_j\left(\opcdag{j}-\opc{j}\right)\,,\quad{\rm with}\quad \hat{\tau}_j\equiv\prod_{l<j}\hat{\sigma}_{l}^x\,,
\end{eqnarray}
we can write the Hamiltonian in Eq.~\eqref{h11} as a quadratic fermionic form
\begin{equation} \label{hamor}
 \hat{H}(t) = 
%\sum_{i,j=1}^{2L} \hat{\Psi}_i^\dagger \widetilde{H}_{ij}(t) \hat{\Psi}_j 
\hat{\mathbf{\Psi}}^{\dagger} \cdot {\mathbb H}(t) \cdot \hat{\mathbf{\Psi}}  
  = \left( \begin{array}{cc}  \opbfcdag{} & \opbfc{} \end{array} \right)
  \left( \begin{array}{rr} {\bf A}(t) & {\bf B}(t) \\
                                        -{\bf B}(t) & -{\bf A}(t) \end{array} \right)
                                        \left( \begin{array}{l}  \opbfc{} \\ \opbfcdag{} \end{array} \right)
                                        \;.
\end{equation}
Here $\hat{\mathbf{\Psi}}$ are $2L$-components (Nambu) fermionic operators defined as 
$\hat{\Psi}_j = \hat{c}_j$  (for $1\le j\le L$) and $\hat{\Psi}_{L+j} = \hat{c}_j^\dagger$,  
and  ${\mathbb H}$ is a $2L\times 2L$ Hermitian matrix having the explicit form shown on the right-hand side, 
with $\bf A$ an $L\times L$ real symmetric matrix, $\bf B$ an $L\times L$ real anti-symmetric matrix. 
Such a form of ${\mathbb H}$ implies a particle-hole symmetry: if $({\bf u}_{\alpha}, {\bf v}_{\beta})^T$ is an instantaneous eigenvector 
of ${\mathbb H}$ with eigenvalue $\epsilon_{\beta}\ge 0$, then $({\bf v}_{\beta}^*, {\bf u}_{\beta}^*)^T$ is an eigenvector with 
eigenvalue $-\epsilon_{\beta}\le 0$.

Let us now focus on a given time, $t=0$, or alternatively suppose that the Hamiltonian is time-independent. 
Then, we can apply a unitary Bogoliubov transformation  
\begin{equation} \label{trasgro}
  \hat{\mathbf{\Psi}} = \left( \begin{array}{l}  \opbfc{} \\ \opbfcdag{} \end{array} \right) = 
  {\mathbb U}_0 \cdot \left( \begin{array}{l} \opbfgamma{} \\ \opbfgammadag{} \end{array} \right) =
  \left(\begin{array}{rr} {\bf U}_0 & {\bf V}^*_0 \\
                                       {\bf V}_0 & {\bf U}^*_0 \end{array} \right) \cdot  
                                       \left( \begin{array}{l} \opbfgamma{} \\ \opbfgammadag{} \end{array} \right) \;,
\end{equation}
where ${\bf U}_0$ and ${\bf V}_0$ are $L\times L$ matrices collecting all the eigenvectors of $\mathbb H$, by column, turning the Hamiltonian 
in Eq.~\eqref{hamor} in the diagonal form
\begin{equation}
  \hat{H} = \sum_{\beta=1}^L \frac{\epsilon_\beta}{2} \left(\opgammadag{\beta} \opgamma{\beta} - \opgamma{\beta}\opgammadag{\beta}\right)\,,
\end{equation}
where the $\opgamma{\beta}$ are new quasi-particle Fermionic operators. 
%and the Nambu spinor $\hat{\Phi}$ is defined in terms of them as 
%$\Phi_{\alpha} = \gamma_{\alpha}$ and $\Phi_{L+\alpha} = \gamma_{\alpha}^\dagger$ for $1\le \alpha\le L$. 
%
%Therefore, also in the inhomogenous case the system can be described in terms of Fermionic quasi-particles, we will see soon that this will be true not only in the static but also in the dynamic case.
%
%The ground state $\ket{{\rm GS}}$ has energy $E_{\rm GS}=-\sum_\alpha \epsilon_\alpha$ and is the vacuum of the $\opgamma{\alpha}$ for all values of 
%$\alpha$: $\opgamma{\alpha}\ket{{\rm GS}}=0$. 
%and, consistently with the Fermionic commutation rules $\bra{{\rm GS}(0)}\gamma_{d\,\alpha}(0) \gamma_{d\,\alpha}(0)^\dagger\ket{{\rm GS}(0)}=1$
%\footnote{We notice that it would be easy to implement a coherent evolution of a system initially in thermal equilibrium at temperature 
%$T=1/(k_B\beta)$, by imposing at time $t=0$ that $\mean{\gamma_{\alpha}^\dagger \gamma_{\alpha}}_0=\frac{1}{\nep^{\beta \epsilon_\alpha}+1}$ 
%and going on with the following analysis.}

To discuss the quantum dynamics when ${\hat H}(t)$ depends on time, we start by writing the Heisenberg equations of
motion for the $\hat{\mathbf{\Psi}}$: they are {\em linear}, due to the quadratic nature of $\hat{H}(t)$.
A simple calculation shows that:
\begin{equation}
i\hbar \frac{d}{dt} \hat{\mathbf{\Psi}}_{H}(t) = 2 {\mathbb H}(t) \cdot \hat{\mathbf{\Psi}}_{H}(t) \;,
\end{equation}
the factor $2$ on the right-hand side originating from the off-diagonal contributions due to $\{\Psi_j,\Psi_{L+j}\}=1$.
These Heisenberg equations should be solved with the initial condition that, at time $t=0$, is
\begin{equation} \label{trasgro_0}
  \hat{\mathbf{\Psi}}_H(t=0) = \hat{\mathbf{\Psi}} = 
%    \left(\begin{array}{cc} {\bf U}_0 & -{\bf V}^*_0 \\
%                                       {\bf V}_0 & {\bf U}^*_0 \end{array} \right)  
{\mathbb U}_0 \cdot  \left( \begin{array}{l} \opbfgamma{} \\ \opbfgammadag{} \end{array} \right) \;.
\end{equation}
A solution is evidently given by 
\begin{equation} \label{Psi-Heis:eqn}
  \hat{\mathbf{\Psi}}_H(t) = 
%  \left(\begin{array}{cc} {\bf U}(t) & -{\bf V}^*(t)\\
%                                       {\bf V}(t) & {\bf U}^*(t) \end{array} \right)  
{\mathbb U}(t) \cdot  \left( \begin{array}{l} \opbfgamma{} \\ \opbfgammadag{} \end{array} \right) =
\left(\begin{array}{rr} {\bf U}(t) & {\bf V}^*(t) \\
                                       {\bf V}(t) & {\bf U}^*(t) \end{array} \right) \cdot  
                                       \left( \begin{array}{l} \opbfgamma{} \\ \opbfgammadag{} \end{array} \right)
\end{equation}
with the same $\opbfgamma{}$ used to diagonalize the initial $t=0$ problem, as long as the time-dependent coefficients
of the unitary $2L\times2L$ matrix ${\mathbb U}(t)$ satisfy the ordinary linear Bogoliubov-de Gennes time-dependent equations
\begin{equation} \label{bog}
i\hbar \frac{d}{dt} {\mathbb U}(t) 
%\left( \begin{array}{cc} {\bf U}(t)  & -{\bf V}^*(t) \\ {\bf V}(t)  & {\bf U}^*(t) \end{array} \right) 
= 2{\mathbb H} (t) \cdot 
%\left( \begin{array}{cc} {\bf U}(t)  & -{\bf V}^*(t) \\ {\bf V}(t)  & {\bf U}^*(t) \end{array} \right)
{\mathbb U}(t) 
\end{equation}
with initial conditions ${\mathbb U}(t=0)={\mathbb U}_0$. 
%${\bf U}(t=0)={\bf U}_0$ and ${\bf V}(t=0)={\bf V}_0$.
It is easy to verify that the time-dependent Bogoliubov-de Gennes form implies that the operators $\opgamma{\beta}(t)$ in 
the Schr\"odinger picture are time-dependent and annihilate the time-dependent state $|\psi(t)\rangle$. Very interestingly, the operators $\opgamma{\beta}(t)$
in the Heisenberg representation are constant in time.
Notice that ${\mathbb U}(t)$ looks like the unitary evolution operator of a $2L$-dimensional problem with Hamiltonian $2{\mathbb H}(t)$.
This implies that we can use a Floquet analysis to get ${\mathbb U}(t)$ whenever ${\mathbb H}(t)$ is time-periodic. 
If we consider one column of Eq.~\eqref{bog}
\begin{equation}
i\hbar \frac{d}{dt} \left(\begin{array}{c}{\bf u}_{\mu}(t)\\{\bf v}_{\mu}(t)\end{array}\right)
= 2{\mathbb H} (t) \cdot 
\left(\begin{array}{c}{\bf u}_{\mu}(t)\\{\bf v}_{\mu}(t)\end{array}\right)
\end{equation}

(here $\left(\begin{array}{c}{\bf u}_{\mu}(t)\\{\bf v}_{\mu}(t)\end{array}\right)$ is a $2L$-column vector), we can find $2L$ independent Floquet
solutions which -- thanks to the particle-hole symmetry -- appear in pairs
\begin{equation}
 \left(\begin{array}{c}{\bf u}_{P\,\alpha}(t)\\{\bf v}_{P\,\alpha}(t)\end{array}\right)\nep^{-i\mu_\alpha t},\quad 
   \left(\begin{array}{c}{\bf v}_{P\,\alpha}^*(t)\\{\bf u}_{P\,\alpha}^*(t)\end{array}\right)\nep^{i\mu_\alpha t}
\end{equation}
where $\mu_\alpha$ is real and the vectors ${\bf u}_{P\,\alpha}(t)$, ${\bf v}_{P\,\alpha}(t)$ $\tau$-periodic in time.
With these column vectors (we define their elements as $U_{P\,j\alpha}$ and $V_{P\,j\alpha}$) and the phase factors $\nep^{-i\mu_\alpha t}$
it is possible to construct a unitary Bogoliubov transformation analogous to the one in Eq.~\eqref{trasgro}; applying its inverse to
the vector of the initial fermionic operators $\left( \begin{array}{l}  \opbfc{} \\ \opbfcdag{} \end{array} \right)$, we find the new fermionic operators
\begin{equation} \label{operazzi:eqn}
  \opgamma{F\,\alpha}(t) = \sum_{j=1}^L\left[{U}_{P\,j\alpha}^*(t)\opc{j}+{V}_{P\,j\alpha}^*(t)\opcdag{j}\right]\nep^{i\mu_\alpha t}\quad \alpha = 1,\ldots,L
\end{equation}
These operators are $\tau$-periodic up to a phase, so we can define them as ``Floquet operators''. Their evolution plus the unitary transformation
which connects them to the operators $\opc{j}$ completely defines the dynamics of the system. Being the Hamiltonian quadratic, Wick's theorem
applies: the expectations of all operators (and the value of the entanglement entropy~\cite{angelo_arXiv16}) can be written in terms of the two-point 
correlators $\mean{\opc{i}\opc{j}}_t$, $\mean{\opcdag{i}\opc{j}}_t$, which can be expressed in terms of the ${U}_{P\,j\alpha}(t)$, ${V}_{P\,j\alpha}(t)$
and $\mu_\alpha$. For instance, we have
{\small
\begin{eqnarray}
  \mean{\opc{i}\opc{j}}_t &=& \sum_{\alpha,\beta}\big[{U}_{P\,i\alpha}^*(t){U}_{P\,j\beta}^*(t)\mean{\opgamma{F\,\alpha}(0)\opgamma{F\,\beta}(0)}_0\nep^{i(\mu_\alpha+\mu_\beta)t}\nonumber\\
     &+&{U}_{P\,i\alpha}^*(t){V}_{P\,j\beta}(t)\mean{\opgamma{F\,\alpha}(0)\opgammadag{F\,\beta}(0)}_0\nep^{i(\mu_\alpha-\mu_\beta)t}\nonumber\\
     &+&{V}_{P\,i\alpha}(t){U}_{P\,j\beta}^*(t)\mean{\opgammadag{F\,\alpha}(0)\opgamma{F\,\beta}(0)}_0\nep^{i(-\mu_\alpha+\mu_\beta)t}\nonumber\\
     &+&[{V}_{P\,i\alpha}(t){V}_{P\,j\beta}(t)\mean{\opgammadag{F\,\alpha}(0)\opgammadag{F\,\beta}(0)}_0\nep^{-i(\mu_\alpha+\mu_\beta)t}\big]\nonumber\\
\end{eqnarray}
}
where the expectation values at time 0 are lengthy expressions involving ${U}_{P\,j\alpha}(0)$, ${V}_{P\,j\alpha}(0)$ and the elements ${U}^0_{j\mu}(0)$, ${V}^0_{j\mu}(0)$ of the unitary transformation diagonalizing the Hamiltonian at time 0 in Eq.~\eqref{trasgro} (these ones give information on the initial state). 
Therefore, knowing the dynamics of the $\opgamma{F\,\alpha}(t)$, we know everything about the evolution of the system. In particular, we know everything about its
Floquet Hamiltonian. To find it, we notice that the stroboscopic dynamics of each $\opgamma{F\,\alpha}(t)$ is just the multiplication by a phase factor
\begin{equation}
  \opgamma{F\,\alpha}(\delta t+n\tau)=\nep^{i\mu_\alpha n\tau}\opgamma{F\,\alpha}(\delta t)\,.
\end{equation}
On the opposite, in the Heisenberg representation, these operators -- like the operators $\opgamma{\beta}(t)$ of Eq.~\eqref{trasgro_0} -- are constant
\begin{equation}
  \opgamma{F\,\alpha}(\delta t)=\nep^{i\hat{H}_F(\delta t) n\tau}\opgamma{F\,\alpha}(\delta t)\nep^{-i\hat{H}_F(\delta t) n\tau}\,.
\end{equation}
this is possible if the Floquet Hamiltonian has the quadratic fermion form
\begin{equation}
  \hat{H}_F(\delta t) = \sum_{\alpha=1}^L \frac{\mu_\alpha}{2}\left(\opgammadag{F,\alpha}(\delta t)\opgamma{F,\alpha}(\delta t)-\opgamma{F,\alpha}(\delta t)\opgammadag{F,\alpha}(\delta t)\right)\,.
\end{equation}
plus some immaterial constant. We are sure that there are no other terms: from one side the Hamiltonian is quadratic, and also the Floquet Hamiltonian 
needs to be quadratic (otherwise, the Wick's theorem would not be valid at all times); from the other side this Floquet Hamiltonian completely describes the evolution
of all the $\opc{j}$ operators and then the dynamics of all the observables. 

To find the values of the $\mu_\alpha$, we numerically diagonalize the
solution ${\mathbb U}(\delta t+\tau,\delta t)$ of Eq.~\eqref{bog} taking as initial value the identity. We do the simultaneous diagonalization of the
two commuting Hermitian operators 
%
%\begin{eqnarray}
$(1/2)({\mathbb U}(\delta t+\tau,\delta t)+{\mathbb U}^\dagger(\delta t+\tau,\delta t))$ and %\nonumber\\
$(1/2i)({\mathbb U}(\delta t+\tau,\delta t)-{\mathbb U}^\dagger(\delta t+\tau,\delta t))$. 
%\end{eqnarray}
%
In this way we obtain the eigenvalues $\cos(2\mu_\alpha\tau)$ and $\sin(2\mu_\alpha\tau)$, from which we can construct $\nep^{-i\mu_\alpha \tau}$
and $\nep^{i\mu_\alpha \tau}$. The corresponding eigenvectors respectively give the amplitudes of $\opgamma{F\,\alpha}(t)$ and $\opgammadag{F\,\alpha}(t)$ 
(see Eq.~\eqref{operazzi:eqn}).  (We are indebted with G.~E. Santoro for the subroutine performing this diagonalization
operation.) %In this way one obtains the eigenvalues $\cos(2\mu_\alpha\tau)$ and $\sin(2\mu_\alpha\tau)$; due to the periodicity of the sine and the cosine the


\begin{thebibliography}{100}
\bibitem{Noteb}
As we discuss in detail below, for concreteness we consider a step-wise
  driving, with a magnetic field jumping between $h_0+A$ and $h_0-A$.
\bibitem{Note_ex}
By the way what is the lower and what is the upper band is not univocally defined: it depends on the choice of the Brillouin zone. With a different choice, the ground state of the Floquet Hamiltonian would be {$\prod_{k>0}^{\rm ABC} \ket{\phi_k^{-}(0)}$}. This state shows exactly the same phase transitions of the Floquet Ground state (and is the adiabatic continuation of the mostly excited state of the static model), so this point is not very crucial.
\bibitem{Nota_scal}
For the sake of brevity the finite-size analysis of the string order, which
  closely resemble Fig.~\ref{fs_scaling:fig} are not shown here.
\bibitem{Nota_closing}
The number of edge states can change only at the resonance points where also
  the bulk gap closes. If we change the frequency and stay inside the same
  gapped phase, the topological properties of the system do not change and we
  will find the same number of edge states at the interface with the vacuum.
\bibitem{Nota_infty}
In the limit of $t_f\to\infty$, we would obtain a periodic driving with
  $t_n=n\tau+3\tau/4$.
\bibitem{note_roy}
Formulae similar to this and to Eq.~\eqref{UktF2:eqn} can be found also in
  Ref.~\ocite{Roy_arXiv16}.
\end{thebibliography}
\end{document}